# Correcting heterogeneous diagnostic bias when developing clinical prediction models using causal hidden Markov models

Jose Benitez-Aurioles[1], Ricardo Silva[2], Brian McMillan[3], Matthew Sperrin[1]

[1] Division of Informatics, Imaging & Data Sciences, University of Manchester, Manchester, United Kingdom

[2] Department of Statistical Science, UCL, London, United Kingdom

[3] Division of Population Health, Health Services Research & Primary Care, University of Manchester, Manchester, United Kingdom

## Abstract

In routine care, individuals identified a priori as high-risk are usually tested for conditions more frequently. Protected attributes, such as sex, ethnicity, or socioeconomic status, may also determine testing frequency. Such heterogeneous detection rates across a population induce label error. This causes systematic model error for specific groups and biases performance metrics during validation, inducing potential fairness concerns.

This paper proposes a method to correct for such bias in prediction models due to differential diagnostic delay. We use a causal inference framework to define our target estimand: an individual's diagnosis probability in a counterfactual scenario where their diagnosis rate matches that of a reference group. We model the longitudinal process as a hidden Markov model, in which confirmatory test results, whether negative or positive, are emissions from a latent progressive disease stage. We validate our approach in simulated data and apply it to a case study of chronic kidney disease prediction using electronic health records.

In simulations, our method reduces prediction bias and improves calibration-in-the-large, correcting the Observed:Expected ratio in the underdiagnosed group from 1.34 (standard deviation: 0.09) in a model developed without any correction for underdiagnosis bias to 1.02 (0.09). Violations of assumptions in the simulation affected the estimation of model parameters, but the proposed approach nonetheless remained better calibrated than the standard model. In the clinical case study, we identify diabetes as the main driver of observability, with an odds ratio of 10.36 (95% confidence interval, 9.80 – 11.02) in 6-month urine albumin-creatinine ratio testing rate. Using our approach to predict the counterfactual diagnostic rate in patients without diabetes, we improved the Observed:Expected ratio of a developed clinical prediction model from 1.55 (1.51 – 1.59) to 1.01 (0.98 – 1.04).

The proposed approach has potential to reduce bias and unfairness in prediction of future conditions.

## Introduction

Clinical prediction models are often developed or validated using routine clinical data [1]. Compared with prospective clinical studies, these usually have larger sample sizes and are more representative of the general population, but data quality is lower [2,3]. In particular, some clinical outcomes are missed, caught late, or not recorded, and will thus not appear in routine data at the appropriate time [4,5]. In England, for example, around 30% of people with type 2 diabetes or hypertension are undiagnosed [6,7], and models trained on data from clinical practice will underestimate the overall incidence of these conditions.

Diagnostic heterogeneity occurs when a patient's characteristics affect their likelihood of diagnosis. These characteristics can be well-known risk factors, as patients viewed as higher risk will usually be monitored more closely, or protected attributes like sex, ethnicity, or socioeconomic status, which might affect how a patient and a healthcare system interact. Models trained on routine clinical care data will thus comparatively underestimate some groups' overall risk, making these groups less likely to be screened or to receive preventive treatment than they should [8,9]. This has been observed across many medical conditions, such as chronic obstructive pulmonary disease [10], cancer [11], and chronic kidney disease [12]. Diagnostic heterogeneity is hard to address, as it is not easily measured from observable data. A particularly difficult question is how to leverage the predictive ability of such attributes, which can benefit some groups [13], without carrying over the bias due to the group's relative underdiagnosis compared with the rest of the population [14,15].

Modellers can often access auxiliary data beyond core predictors and outcomes. This includes records of clinical interactions, such as diagnostic tests, and point-of-diagnosis information, such as biomarker values or symptom severity. While the problem of differential underdiagnosis is unsolvable with only core predictors and outcomes, the additional information is currently used to understand whether some patients are underserved by current diagnostic practice [16]. In cancer epidemiology, for example, stage at diagnosis is used as a marker to study diagnostic delay in cancer, as it is understood that groups diagnosed at later stages of the disease are underdiagnosed at earlier stages [11]. The underlying reasoning in progressive diseases (i.e., conditions which, if untreated, worsen over time) is that patients remain undiagnosed until symptoms or clinical signs become severe enough for the condition to be detected by the healthcare system. While this has been used to identify potential diagnostic heterogeneity, it has not, to our knowledge, been used to correct for it in clinical prediction models.

This work proposes a methodology that leverages these diagnostic markers to correct for differential bias due to underdiagnosis in clinical prediction models for progressive diseases. We take a causal approach, modelling the problem through a directed acyclic graph (DAG) in which the probability of receiving a diagnostic test is affected by both the disease stage and the observability attributes (i.e., attributes that influence an individual's probability of undergoing diagnostic testing conditional on disease stage, while also potentially affecting disease risk). Our objective is to estimate a counterfactual outcome distribution where we intervene on the diagnostic process, not the patient's attributes or the disease process. Specifically, we estimate an individual's risk as if they were subject to the same testing rate as a predefined reference group, chosen to be the group with the smallest diagnostic delays. We show that this can be estimated by treating the unmeasured disease stage as a hidden Markov model (HMM). A predictive model fit on data imputed by such a HMM will correct for diagnostic heterogeneity bias. We demonstrate the method's validity theoretically and show its identifiability through simulation. We showcase its use in a real-world example predicting chronic kidney disease using UK electronic health records.

## Methods

### Notation

The directed acyclic graph (DAG) of Figure 1a summarises the variables considered and structural assumptions made. Consider a set of $N$ individuals, where $n = 1, .., N$ is an individual index, with predictor vectors $X_n$, measured at baseline, and observability attributes $A_n$, which are predictors that influence the diagnostic process, in particular the probability of undergoing testing or screening. These may include demographic characteristics or clinical factors that affect surveillance intensity. To keep the derivation clear, we consider $A_n$ to be binary, with $A_n = 0$ denoting a reference (well-monitored) group, and $A_n = 1$ a relatively underdiagnosed one. We define $\tilde{A}_n$ as the presentation of the observability attribute, i.e., the component of $A_n$ that directly enters the testing decision conditional on disease stage. For example, if $A_n$ is the presence of a comorbidity, then $\tilde{A}_n$ represents the influence of the comorbidity on surveillance for the target clinical condition. Thus, disease risk may depend on $A_n$, but the probability of being tested given disease stage depends on $\tilde{A}_n$. In observed data, $A_n$ and $\tilde{A}_n$ are always the same, and this distinction is only introduced to define a counterfactual intervention on the diagnostic process in a later section. Each individual is followed for $T_n \leq T$ timepoints. At each discrete time after baseline $t \in \{1, .., T_n\}$, an individual $n$ is characterised by their disease stage $S_{n,t} \in \{0,1,2\}$, where 0 denotes no disease, 1 an early or less advanced stage, and 2 a later stage. In addition, test results at each $t$ are recorded as $R_{n,t} \in \{0,1,2,3\}$, corresponding to confirmatory negative, early stage positive, late stage positive, or no confirmatory test at that timepoint respectively. Individuals are diagnosed, $D_{n,t} = 1$, when they are tested and the result of the test is positive. After diagnosis, individuals are censored and no longer followed.

### Assumptions

We assume disease incidence is determined by a hazard function $h(X_n, A_n, t)$ which depends on baseline covariates $X_n$ and $A_n$, as well as time $t$. All individuals without the disease may progress to early stage $S_{n,t} = 1$, and then late stage $S_{n,t} = 2$ afterwards, where we assume that early-to-late progression, $P(S_{n,t} = 2|S_{n,t-1} = 1, X_n, A_n)$, follows a common stochastic process so that the probability of transitioning from early to late stage is the same for all timepoints, depending only on the immediately preceding disease stage. The disease is progressive, so that individuals can progress from having no disease to early, and from early to late, but not regress to an earlier stage.

We assume that testing rates are constant through time, conditional on current disease stage and observability attribute $A_n$ (while being independent of $X_n$), so that groups with $A_n = 0$ and $A_n = 1$ are tested at different rates and therefore experience different levels of delay in diagnosis. Finally, we consider the tests $R_{n,t} = \{0,1,2\}$ to have perfect sensitivity and specificity, so that they confirm, at that timepoint, the stage of the individual $S_{n,t} = \{0,1,2\}$.

We distinguish between assumptions required to define the model and counterfactual estimands below, detailed above and in the directed acyclic graph of Figure 1a, and assumptions introduced to simplify the derivation and notation, as well as to aid convergence. We make two additional assumptions of the latter type. First, we assume that early-to-late disease progression is independent of baseline covariates, so that $P\left(S_{n,t} = 2\middle|S_{n,t-1} = 1, X_n, A_n\right) = P\left(S_{n,t} = 2\middle|S_{n,t-1} = 1\right)$. Second, we set the testing rate in late-stage disease to be independent of $A_n$, as for late-stage cases, symptoms, biomarkers, or signs of the disease are assumed to be sufficiently severe to prompt diagnosis regardless of observability differences. We discuss the implications and potential relaxation of these assumptions in the discussion.

As individuals who are diagnosed are subsequently censored, we do not assume anything about disease progression or monitoring in early stage or late stage individuals beyond the point of diagnosis.

## Modelling stage transition and emissions through a hidden Markov chain

This longitudinal process laid out by the assumptions above follows that of a Markov chain, specified by three matrices: the transition kernel $\mathrm{Q}(X_n, A_n, t)$, the emission probabilities $\Gamma(\mathrm{A_n})$, and the initial state $\pi(X_n, A_n)$.

The $3 \times 3$-sized transition kernel $\mathrm{Q}(X_n, A_n, t)$ consists of the elements $\mathrm{Q_{ij}}(X_n, A_n, t)$ which define the probability of an individual to transition from stage $S_{n,t-1} = i$ to $S_{n,t} = j$. We have:

$$\mathrm{Q_{ij}}(X_n, A_n, t) = P\left(S_{n,t} = j \middle| X_n, A_n, S_{n,t-1} = i\right) = \begin{pmatrix} 1 - h(X_n, A_n, t) & h(X_n, A_n, t) & 0 \\ 0 & 1 - \beta^{S=1\to2} & \beta^{S=1\to2} \\ 0 & 0 & 1 \end{pmatrix} \quad (1)$$

Here, $h(X_n, A_n, t)$ is a (parametric or non-parametric) hazard function, and $\beta^{S=1\to2}$ is the constant probability of progressing from early to late stage for all individuals. The $3 \times 4$-sized emission probabilities $\Gamma(\mathrm{A_n})$ consists of the elements $\Gamma_{\mathrm{ij}}(\mathrm{A_n})$ which define the probability that an individual with stage $S_{n,t} = i$ gets a testing result $R_{n,t} = j$. We have:

$$\Gamma_{\mathrm{ij}}(A_n) = P\left(R_{n,t} = j \middle| A_n, S_{n,t} = i\right) = \begin{pmatrix} \beta^R_{S=0,A_n} & 0 & 0 & 1 - \beta^R_{S=0,A_n} \\ 0 & \beta^R_{S=1,A_n} & 0 & 1 - \beta^R_{S=1,A_n} \\ 0 & 0 & \beta^R_{S=2} & 1 - \beta^R_{S=2} \end{pmatrix} \quad (2)$$

Here, $\beta^R_{S=0,A_n}$ and $\beta^R_{S=1,A_n}$ are the testing probabilities of individuals without disease or at early stage, which depend on $A_n$ so that $\beta^R_{S=0,A_n=0}$ is not necessarily equal to $\beta^R_{S=0,A_n=1}$ and $\beta^R_{S=1,A_n=0}$ is not necessarily equal to $\beta^R_{S=1,A_n=1}$. $\beta^R_{S=2}$ is the constant testing rate in late stage cases. Finally, the 3-element initial state $\pi(X_n, A_n)$ consists of $\pi_i(X_n, A_n)$ which define the probability that an individual had stage $S_{n,1} = i$ at timepoint $t = 1$. We have:

$$\pi_i(X_n, A_n) = P\left(S_{n,1} = i \middle| X_n, A_n\right) = \begin{pmatrix} 1 - h(X_n, A_n, 1) \\ (1 - \beta^{S_1=2}) h(X_n, A_n, 1) \\ \beta^{S_1=2} h(X_n, A_n, 1) \end{pmatrix} \quad (3)$$

Here, $h(X_n, A_n, 1)$ is the hazard function at the first timepoint. Notably, we do not assume that all individuals at the first timepoint are early stage, so as to account for undiagnosed individuals at baseline, but do consider that the probability of being late stage conditioned on having the disease, $P(S_{n,1} = 2 | S_{n,1} \neq 0)$, is a constant.

This HMM can be estimated using a maximum likelihood approach [17]. Consider a set of observations $\left\{x_n, a_n, \left\{r_{n,t}, d_{n,t}\right\}_{t=1..T_n}\right\}_{n=1..N}$. The objective is to find the parameter set $\theta \in \Theta$ which maximises the likelihood $\prod_{n=1..N} P(R_{n,1..T_n} = r_{n,1..T_n}, D_{n,1..T_n} = d_{n,1..T_n} | X_n = x_n, A_n = a_n; \theta)$. Note that, since $D_{n,t}$ is fully determined by $R_{n,t}$ (as $D_{n,t} = 1$ if $R_{n,t} = 1$ or $R_{n,t} = 2$), it does not need to be included in the likelihood function, so that we instead consider $\prod_{n=1..N} P(R_{n,1..T_n} = r_{n,1..T_n} | X_n = x_n, A_n = a_n; \theta)$.

The likelihood function can be written in terms of the parameters of the HMM, $\theta = \{\pi, \mathrm{Q}, \Gamma\}$ by marginalising over the possible latent states as:

$$P\left(R_{n,1..T_n} \middle| X_n, A_n; \theta\right) = \sum_{S_{n,1..T_n}} P\left(R_{n,1..T_n} \middle| X_n, A_n, S_{n,1..T_n}; \theta\right) \times P\left(S_{n,1..T_n} \middle| X_n, A_n; \theta\right)$$
$$= \sum_{S_{n,1..T_n}} \left( \prod_{t=1..T_n} \Gamma_{S_{n,t}, R_{n,t}} \right) \times \left( \pi_{S_{n,1}} \prod_{t=2..T_n} Q_{S_{n,t-1}, S_{n,t}} \right) \quad (4)$$

While a full expression of the likelihood is possible as above, a more feasible optimisation is to find the choice of parameters $\hat{\theta}$ which maximises the log-likelihood as:

$$\hat{\theta} = argmax_{\theta \in \Theta} \sum_{n=1..N} \log P\left(R_{n,1..T_n} = r_{n,1..T_n} \middle| X = x_n, A = a_n; \theta\right)$$
$$= argmax_{\theta \in \Theta} \sum_{n=1..N} \sum_{t=1..T_n} \log P\left(R_{n,t} = r_{n,t} \middle| X = x_n, A = a_n, R_{n,1..t-1} = r_{n,1..t-1}; \theta\right) \quad (5)$$

The probability of an individual's test result, conditioned on baseline characteristics and past testing history, $P\left(R_{n,t} = r_{n,t} \middle| X = x_n, A = a_n, R_{n,1..t-1} = r_{n,1..t-1}; \theta\right)$, can be derived using a forward algorithm as detailed in Supplementary 1.1. If the chosen hazard function is differentiable, these probabilities will also be differentiable, and maximum likelihood optimisation can be performed using gradient-based methods. For this, we use the Limited-memory Broyden–Fletcher–Goldfarb–Shanno algorithm [18], implemented in PyTorch [19].

## Defining a fair counterfactual estimand

In a predictive problem, we usually want to develop a model which estimates the risk that an individual has of being diagnosed by the prediction horizon $T$, $P(D_{n,T} = 1 | X_n, A_n)$. However, if the underlying process is as described above, this model will prioritise individuals of the reference group $A_n = 0$, when their probability of being diagnosed at early stage $\beta^R_{S=1,A=0}$ is higher than that of the underserved group $A_n = 1$, $\beta^R_{S=1,A=1}$.

We define an intervention on the presentation of the observability attribute $\tilde{A}_n$, as laid out in Figure 1b. Our new estimand is $P(D_{n,T}^{\tilde{A}_n=0} = 1 | X_n, A_n)$, the counterfactual probability that an individual would be diagnosed had they been presented as part of the reference group, $\widetilde{A_n} = 0$. Although this intervention is not necessarily well-defined, it can be interpreted as an intervention on clinical decision-making, making the healthcare system behave as if the individual belonged to the reference group when determining whether they should be tested or not, conditional on disease stage.

A model targeting this new estimand ensures that an individual's score does not depend on differences in the testing rate across groups and only considers genuine differences of risk, whether biological, social, or otherwise. Similarly to path-specific interventions [20], we thus want to conserve the effect of $A_n$ on $D_{n,t}$ through $S_{n,t}$, while blocking the effect of $A_n$ on $D_{n,t}$ through the direct path $A_n \rightarrow R_{n,t}$.

## Using HMM-imputed data to train an unbiased counterfactual model

Once the HMM is fitted, it can be used to provide a probabilistic reconstruction of each individual's unobserved disease trajectory over time, given observed data, $P(S_{n,t} | X_n = x_n, A_n = a_n, R_{1..T_n} = r_{1..T_n})$. Using this estimate and the fitted model, we can also estimate the counterfactual probability that the individual would be diagnosed by the end of follow-up under the $\tilde{A}_n = 0$ intervention, $P(D_{n,T}^{\tilde{A}_n=0} = 1 | X_n = x_n, A_n = a_n, R_{1..T_n} = r_{1..T_n})$. This quantity represents the individual-level risk of diagnosis under the reference group's testing rate and is our target for outcome re-imputation (see details on how to derive these estimates in Supplementary 1.2).

For each individual in the data, we can thus obtain the counterfactual probability of having been diagnosed, given their testing history:

$$\hat{p}_{n,T}^{cf} = P\left(D_{n,T}^{\tilde{A}_n=0} = 1 \middle| X_n = x_n, A_n = a_n, R_{1..T_n} = r_{1..T_n}\right) \quad (6)$$

The objective is to use these probabilities $\hat{p}_{n,T}^{cf}$ to construct a counterfactual outcome $d_{n,T}^{cf}$ for each individual in the dataset, so that a model trained or validated to predict $d_{n,T}^{cf}$ is less biased with respect to the estimand $P(D_{n,T}^{\tilde{A}_n=0} = 1 | X_n, A_n)$.

In theory, re-imputing the entire dataset so that $d_{n,T}^{cf} \sim Bernoulli(\hat{p}_{n,T}^{cf})$ should lead to an unbiased estimate. However, if the hazard model or HMM are misspecified, this will propagate those misspecifications throughout the re-imputed data. A more conservative approach is to impute as few individuals as possible. Firstly, due to how in observed data $A_n$ and $\tilde{A}_n$ are always the same, $P\left(D_{n,T}^{\tilde{A}_n=0} = 1 \middle| X_n = x_n, A_n = 0, R_{1..T_n} = r_{1..T_n}\right) = P\left(D_{n,T}^{\tilde{A}_n=0} = 1 \middle| X_n = x_n, A_n = 0, \tilde{A}_n = 0, R_{1..T_n} = r_{1..T_n}\right) = P(D_{n,T} = 1 | X_n = x_n, A_n = 0, R_{1..T_n} = r_{1..T_n})$ by consistency. Therefore, no imputation is required for those with $a_n = 0$. Secondly, we assume monotonicity, so that if an individual was diagnosed, and their probability of being diagnosed is higher in the counterfactual scenario, then they remain diagnosed, so that $d_{n,T} = 1$ implies $d_{n,T}^{cf} = 1$. Thus, we only re-impute the outcome, making $d_{n,T}^{cf}$ potentially different from $d_{n,T}$, in individuals with $a_n = 1$ and $d_{n,T} = 0$.

Since we impose monotonicity, such that $d_{n,T} = 1 \Rightarrow d_{n,T}^{cf} = 1$, individuals who are already diagnosed are deterministically assigned to a counterfactual diagnosis, even though their counterfactual diagnosis probability $p_{n,T}^{cf}$ may be smaller than one. If we were to re-impute outcomes for undiagnosed individuals using $p_{n,T}^{cf}$ without adjustment, this would inflate the overall counterfactual incidence in the $a_n = 1$ group, because the realised number of diagnoses would be much higher than the total counterfactual probability mass:

$$\sum_{a_n=1, d_{n,T}=0} p_{n,T}^{cf} + \sum_{a_n=1, d_{n,T}=1} p_{n,T}^{cf} < \sum_{a_n=1, d_{n,T}=0} p_{n,T}^{cf} + \sum_{a_n=1, d_{n,T}=1} 1 \quad (7)$$

To ensure that the re-imputed outcomes are calibrated to the expected counterfactual incidence ($\sum_{n=1..N} d_{n,T}^{cf}$), we rescale the counterfactual probabilities of those which we re-impute. Specifically, we define:

$$\hat{p}_{n,T}^{cf-recalibrated} = \hat{p}_{n,T}^{cf} \times \frac{\bar{p}_{A=1}^{cf} - \bar{d}_{A=1}}{\bar{p}_{A=1,D=0}^{cf}} \quad (8)$$

Where $\bar{p}_{A=1}^{cf}$ is the estimated counterfactual incidence in the $a_n = 1$ group, $\bar{d}_{A=1}$ is the observed incidence, and $\bar{p}_{A=1,D=0}^{cf}$ is the average counterfactual probability among individuals whose outcome is re-imputed. This rescaling preserves monotonicity, so that we still have $d_{n,T} = 1 \rightarrow d_{n,T}^{cf} = 1$, while ensuring that the incidence of the re-imputed dataset corresponds to the overall estimated counterfactual incidence. Finally, a model trained to predict the imputed outcome $d_{n,T}^{cf}$ will be unbiased to the counterfactual estimand $P\left(D_T^{\tilde{A}=0} = 1 \middle| X, A\right)$, as desired.

## Simulation

To show the identifiability of the HMM and validate the approach, we use a simulation. Four scenarios are considered:

**Scenario 1:** We simulate a set of $N = 50{,}000$ patients over $T = 10$ timepoints, with $\Pr(A_n = 1) = 0.2$. Two additional predictors are drawn as:

$$X_{n,1} \sim \mathbb{I}(A = 0) \times \mathcal{N}(0,1^2) + \mathbb{I}(A = 1) \times \mathcal{N}(1,1.5^2), \qquad X_{n,2} \sim \mathcal{N}(0.5,1^2) \quad (9)$$

So that $X_{n,1}$ correlates with $A$. The incidence follows a Weibull hazard function:

$$P\left(S_{n,t} = 1 \middle| X_n = (x_{1,n}, x_{2,n}), A_n = a_n, S_{n,t-1} = 0\right) = 0.005 \times 1.5 \times \left(\frac{t}{T}\right)^{1.5-1} \times \mathrm{e}^{0.5x_{1,n} - 0.25x_{2,n} + 0.25a_n} \quad (10)$$

Cases at $t = 1$ start as early stage, so $\beta^{S_1=2} = P(S_{n,1} = 2 | S_{n,1} \neq 0) = 0$, and the early-to-late progression rate is $\beta^{S=1\rightarrow 2} = 0.1$. Testing rates are $\beta_{S=0,A=0}^{R} = 0.025$, $\beta_{S=0,A=1}^{R} = 0.01$, $\beta_{S=1,A=0}^{R} = 0.1$, $\beta_{S=1,A=1}^{R} = 0.05$, and $\beta_{S=2}^{R} = 0.3$.

In the next three scenarios, we explore the robustness of the approach by violating different assumptions. Otherwise, the conditions are the same as those of scenario 1.

**Scenario 2:** We violate the assumption that tests have perfect sensitivity, by exploring a scenario in which the test has a sensitivity of 90% in early stage cases, and 95% in late stage cases. Thus, an individual with the condition might be tested, but be incorrectly confirmed as negative.

**Scenario 3:** We violate the assumption that disease progression has an equal rate across all individuals, so that now the underserved group progresses faster, with $\beta_{A=0}^{S=1\to2} = 0.1$ and $\beta_{A=1}^{S=1\to2} = 0.13$.

**Scenario 4:** In an open cohort scenario, individuals at baseline will already have diagnostic gaps across groups at baseline. To simulate this more realistic scenario, we run the simulation for $T_{total} = 20$ instead of 10. The baseline is now defined as $t = 10$, so that testing information before that is not available.

In all scenarios, the imputation follows the approach previously described, with the hazard parametrised by a Weibull distribution. Each scenario is run 200 times. We report the average estimate, bias, empirical standard error, and mean squared error, of each fitted parameter in the HMM. In addition, four logistic regression models are trained: 1) a 'Naïve' model, trained to predict $d_{n,T}$ using $x_n$ and $a_n$; 2) a 'Blind' model, trained to predict $d_{n,T}$ using only $x_n$; 3) an 'Imputed' model, trained to predict our approach's imputed $d_{n,T}^{cf}$ using $x_n$ and $a_n$; and 4) an 'Ideal' model, trained on a separate dataset, already sampled from the counterfactual distribution where testing is independent of $A_n$. All models are validated on a new dataset with $N_{val} = 50{,}000$ sampled from the counterfactual distribution. Performance is assessed through the area under the receiver-operator curve (AUROC), the calibration slope and intercept, the Observed:Expected (O:E) ratio, and decile calibration plots. The performance is evaluated overall and in groups $A_n = 0$ and $A_n = 1$ separately, with the average and standard deviation across replications reported.

## Simulation: Results

In Scenario 1, where all assumptions are met, Table 1 shows the performance of the HMM fit across the 200 simulation runs. Most parameters were estimated with low bias, with the largest bias observed in the progression rate, which was overestimated at 0.113 (standard error: 0.047), versus the true value of 0.1. The standard error of the estimated parameters was overall small, with it being the highest in the estimation of the late stage testing rate, estimated on average as 0.302 (0.122), versus the true value of 0.3.

Figure 2 presents calibration plots comparing the HMM-imputed model to the benchmark models. In Scenario 1, calibration in the underserved group ($A_n = 1$) was improved relative to the 'Naïve' and 'Blind' models, with O:E ratios of 1.01 (standard deviation, 0.09), 1.34 (0.09), and 1.30 (0.07), respectively (see Supplementary 3, Table 4 for full performance metrics).

Scenario 2, in which test sensitivity was 90% in early stage and 95% in late stage cases, was the most challenging for the proposed method. Testing rates were substantially overestimated (Supplementary 3, Table 1), from the true values of $\beta_{S=1,A=0}^{R} = 0.1$, $\beta_{S=1,A=1}^{R} = 0.05$, and $\beta_{S=2}^{R} = 0.3$, to estimated averages (standard deviation) of 0.440 (0.116), 0.280 (0.095), and 0.536 (0.159), respectively. The optimisation algorithm particularly overestimated the progression rate, $\beta^{S=1\rightarrow 2} = 0.1$ to 0.650 (0.263). Despite this misspecification, the HMM-imputed model still outperformed the 'Naïve' and 'Blind' models in the underserved group, with O:E ratios of 1.20 (0.09), 1.33 (0.09), and 1.30 (0.07), respectively (Figure 2; Supplementary 3, Table 5).

Our approach was moderately robust in Scenario 3, in which progression rates differed across groups (Supplementary 3, Tables 2 and 6), although the model slightly overpredicted the risk in the underserved group, with an O:E ratio of 0.90 (0.07). Scenario 4 (Supplementary 3, Tables 3 and 7), in which underdiagnosis was present at baseline, was handled well by the method, yielding a well-calibrated model with an O:E ratio of 1.01 (0.05).

## Real-world example: developing a prognostic chronic kidney disease model

We applied the method to develop and validate a prognostic model for chronic kidney disease (CKD), using UK Biobank participants [21] with primary care data linkage. Individuals with diagnosed CKD at baseline (i.e., study entry) were excluded. CKD stage at diagnosis was determined from Read and ICD-10 codes, with stages 1-2 classified as early, and 3-5 as late. Individuals with missing stage at diagnosis were imputed using predictive mean matching based on baseline covariates.

Albumin-to-creatinine ratio tests prior to diagnosis were extracted and coded as confirmatory negative tests if they showed a ratio lower than 3 mg/mmol. Estimated glomerular filtration rate (eGFR) measurements were not considered, as although an eGFR rate lower than 60 mL/min/m$^2$ could lead to a CKD diagnosis by itself, one higher than that does not necessarily rule out the condition. Follow-up was divided into 6-month intervals over 5 years ($T = 10$). Predictors included age, sex, Townsend quintile, BMI, smoking, hypertension, cardiovascular disease (CVD), and diabetes.

Instead of only considering one binary attribute as done in the derivation, we wanted to account for multiple covariates at baseline which, clinically, we knew were likely to contribute to observability and speed of diagnosis: age, diabetes, CVD, hypertension, ethnicity (White vs non-White), sex (male vs female) and deprivation (most deprived Townsend quintile vs others). Because of this, we modified the emission matrix of (2), to be:

$$\Gamma_{\mathrm{ij}}(A_n) = \begin{pmatrix} b(A_n) & 0 & 0 & 1 - b(A_n) \\ 0 & b(A_n) & 0 & 1 - b(A_n) \\ 0 & 0 & \beta^R_{S=2} & 1 - \beta^R_{S=2} \end{pmatrix} \quad (11)$$

Where $\beta^R_{S=2}$ is still the rate of testing in late stage cases, $A_n$ is the vector of observability attributes, and:

$$b(A_n) = expit\big(\beta_{intercept} + \beta_{age} \times Age + \beta_{dia} \times Diabetes + \beta_{cvd} \times CVD + \beta_{hyp} \times Hypertension + \beta_{nwh} \times NonWhite + \beta_{fem} \times Female + \beta_{twn} \times InTopTownsendQuintile\big) \quad (12)$$

Is a parametrised functional form for the testing rate modelled through logistic regression. Here, we chose $b(A_n)$ to be the same for those with no CKD and early stage CKD, as early stage CKD is asymptomatic and unlikely to be detected without urine testing, unlike late stage CKD which may be detected with more routine blood tests. This clinically reasonable addition improved model fit, as without it, strong parameter correlations caused the maximum likelihood algorithm to diverge. The HMM was fitted with this emission matrix, and a Cox Proportional-Hazard model was chosen to model the hazard function $h(X_n, A_n)$, where $X_n$ is the vector for the predictors that do not affect observability (BMI and smoking status).

In this case, we chose the counterfactual intervention to change the observability of all individuals to those with diabetes, hypertension, CVD, who also are White, male, and outside the top quintile of deprivation. This was chosen as the reference group as we considered it to potentially have the shortest diagnostic delays. Thus:

$$b^{cf}(A_n) = expit\big(\beta_{intercept} + \beta_{age} \times Age + \beta_{dia} \times 0 + \beta_{cvd} \times 0 + \beta_{hyp} \times 0 + \beta_{nwh} \times 0 + \beta_{fem} \times 0 + \beta_{twn} \times 0\,\big) \quad (13)$$

And:

$$\Gamma_{\mathrm{ij}}^{cf}(A_n) = \begin{pmatrix} b^{cf}(A_n) & 0 & 0 & 1 - b^{cf}(A_n) \\ 0 & b^{cf}(A_n) & 0 & 1 - b^{cf}(A_n) \\ 0 & 0 & \beta_{S=2}^{R} & 1 - \beta_{S=2}^{R} \end{pmatrix} \quad (14)$$

Three predictive models are validated in a counterfactually imputed dataset, where the HMM was fitted and used to re-sample negative cases in individuals outside of the reference group. The three explored models are: 1) a 'Naïve' logistic regression model, which uses all covariates to predict the outcome as it would normally be done; 2) a 'Blind' logistic regression model, which excludes ethnicity, sex, and Townsend quintile as predictors; and 3) an 'Imputed' regression model, using all available predictors to predict the counterfactually imputed outcome. All three models predict 5-year incidence of CKD. Performance metrics are adjusted for optimism by bootstrapping [22], re-running the HMM imputation in each re-sample. Performance was assessed using AUROC, calibration slope, O:E ratio, logistic error, quintile calibration plots, and net benefit decision curves [23] between thresholds of 0.05 and 0.30. To visualise the proportion of undiagnosed individuals, the proportions of diagnosed ($S_t \neq 0, D_t = 1$) and undiagnosed ($S_t \neq 0, D_t = 0$), as estimated by the HMM, were plotted across timepoints. Further details on data preparation, model fitting, and validation, are provided in Supplementary 2.

## Real-world example: Results

A total of 123,460 individuals were included in the final cohort, with 7,090 cases of CKD observed by the end of follow-up, of which 3,524 occurred within 5 years. Of these, 2.8%, 18.5%, 75.2%, 0.6%, and 0.2% were diagnosed at stages 1–5 respectively, while 2.7% had a CKD diagnosis without a recorded stage. After missing data imputation, 735 individuals were classified as having early stage CKD at diagnosis, and 2789 as having late stage. Additional baseline characteristics are presented in Supplementary 4, Tables 1.

Parameter estimates from the fitted HMMs are shown in Table 1, with those related to the hazard model available in Supplementary 4, Table 2. The estimated parameters of the HMM were consistent with clinical expectations. The early-to-late 6-month progression rate found was 0.125 (95% confidence interval: 0.114 – 0.139), corresponding to an average time of 48 (44 – 52) months to progress from the early to the late stages of CKD. The testing rate for those without CKD or with early stage CKD was low, being, on average, 0.0252 (0.0249 – 0.0257). Having diabetes (coefficient: 2.34, 2.28 – 2.40; odds ratio: 10.36, 9.80 – 11.02) and hypertension (1.02, 0.977 – 1.07; odds ratio: 2.78, 2.65 – 2.92) were the two attributes which were associated with a higher rate of testing. In terms of protected attributes, being female was associated with a lower rate of testing (-0.364, -0.406 – -0.329; odds ratio: 0.69, 0.67 – 0.72), while not being White (0.397, 0.334 – 0.489; odds ratio: 1.49, 1.40 – 1.63) and being in the top quintile of deprivation (0.108, 0.058 – 0.154; odds ratio: 1.11, 1.06 – 1.17) were associated with a higher rate. Testing rate in late stage individuals was much higher, being 0.584 (0.558 – 0.606), meaning that individuals usually stayed undiagnosed during the early stage of the disease, and got diagnosed quickly as soon as they progressed to late stage. Looking at the estimated evolving prevalence of undiagnosed and diagnosed individuals (Figure 3, Supplementary 4 Figures 1 and 2), 58% of individuals were diagnosed by the 5 year end of follow-up, with the most accurately diagnosed being individuals with diabetes, as 80% of those who had CKD and diabetes were estimated to be diagnosed by the end of follow-up.

When evaluating the calibration of models against the counterfactual estimand (Figure 4, with other calibration curves included in Supplementary 4, Figures 3 and 4), the Naïve model underpredicted risk in individuals without diabetes, with an overall O:E ratio of 1.55 (1.51 – 1.59) versus 1.13 (1.03 – 1.24) in those with diabetes. This was the same for the Blind model, with 1.55 (1.52 – 1.59) and 1.13 (1.01 – 1.25) for the two groups. The Imputed model corrected this bias, with an O:E ratio of 1.01 (0.98 – 1.04) and 1.01 (0.91 – 1.09) for the two groups instead (the rest of the performance metrics are presented in Supplementary 4, Tables 3 – 9). This correction produced improved net benefit across both groups, as shown by decision curve analysis (Supplementary 4, Figures 5 – 10). Looking at the coefficients of each of the three final models (Supplementary 4, Table 10), the Diabetes, Hypertension and Cardiovascular disease remained the predictors with the biggest coefficients, being 0.68, 0.59, and 0.43 respectively for the Naïve model, 0.68, 0.59 and 0.41 for the Blind model, and 0.39, 0.56 and 0.39 for the Imputed model.

## Discussion

We proposed a method to correct bias arising from differences in diagnostic rates across protected groups. Unlike existing approaches that depend on external incidence estimates or fixed assumptions, our method relies entirely on observed data. In simulations, the model showed empirical identifiability and improved calibration for underdiagnosed groups, demonstrating robustness even when key assumptions were violated.

Applied to chronic kidney disease (CKD), the estimated early-to-late progression time of about four years was consistent with published studies on disease progression [24]. The finding that around half of patients with CKD were diagnosed by the end of follow-up also aligned with previous UK research [25]. Overall, the method identified systematic differences in testing rates across population subgroups. Groups at higher risk, such as those with diabetes, hypertension, or cardiovascular disease, were tested more often and diagnosed earlier, consistent with clinical expectations for CKD. Being female was the only protected attribute linked to delayed diagnosis, in agreement with reports of higher underdiagnosis in women compared with men [12]. Although we observed apparently counterintuitive longer delays in diagnosis for White and less deprived patients, this could reflect comorbidities which we did not account for, leading clinicians to test non-White and more deprived patients more frequently. If these differences are deemed potentially misleading, the counterfactual definition could be modified to exclude interventions on ethnicity and deprivation. Still, the influence of protected attributes on observability was small compared with that of clinical factors, particularly diabetes. The model trained on the hidden Markov model (HMM)-imputed data improved calibration and clinical utility, increasing estimated risk, especially among patients without diabetes, without reducing it among better-monitored individuals. This change is clinically meaningful: training a prediction model on counterfactually imputed data avoided propagating diagnostic disparities in CKD detection present in routine practice, leading to more accurate risk stratification and patient benefit.

Our approach builds on prior work on the identifiability of HMMs for progressive diseases [26,27] and their use in causal inference to estimate effects on latent variables [28,29] or to address unobserved confounding [30]. However, the application of causal HMMs to address subgroup diagnostic delay bias in clinical prediction has not previously been explored. This study is also connected to the frameworks of counterfactual fairness [31] and path-specific counterfactual fairness [20], which ensure fairness by removing all or selected parts of the influence of a protected attribute on predictions. In our setting, we block the pathway from protected group membership (or other groups defined by observability attributes) to outcomes through testing probability, while retaining its effect on disease development. Related work on label error bias has usually addressed the issue of label error by incorporating external incidence estimates or using anchor points, i.e., cases where the true outcome is known with high confidence [32].

The results of this work should be interpreted in light of the limitations of the model considered. Some assumptions were introduced primarily for fitting convergence and are, in principle, simple to relax. In particular, we assumed a single early-to-late progression rate equal across individuals to avoid issues with strong correlations between disease progression and diagnostic delay parameters, as diagnosis at late-stage can arise either from faster progression or from less frequent monitoring. Relaxing this assumption might introduce practical identifiability challenges. Similarly, one could allow late-stage testing rates to vary across patient characteristics, but we chose to not consider this as it is a reasonable assumption to reduce model variance.

Other assumptions are more structural and require further conceptual development. In particular, the first-order Markov assumption made by the causal diagram may be unrealistic, as disease progression is likely to depend on longer-term patient trajectory. This could be partially addressed by allowing different progression rates across patients as discussed above, or by extending the framework to higher-order Markov models. Finally, the approach could be extended to richer representations of disease trajectory by considering more than two discrete stages or continuous disease markers (e.g., stages I to V in CKD, or diabetes biomarkers such as HbA1c). Similarly, the diagnostic pathway could be considered more wholly by including multiple confirmatory test types, potentially with imperfect sensitivity or specificity, or auxiliary clinical observations that reflect latent disease status without leading to diagnosis.

As fairness becomes increasingly central to predictive modelling, developers must consider how modelling decisions affect patient benefit across the population. Evaluating model performance within contextualised groups of interest [33], defined by protected attributes or clinical factors, is important, but so is recognising that such assessments can themselves be biased due to the inherent issues of real-world clinical data. Recent calls for a “data-centric” approach to fairness [34] emphasise addressing these sources of bias directly. Our method aligns with this perspective by explicitly modelling the data-recording mechanisms that generate bias and defining a counterfactual model that is unbiased with respect to the estimand of interest.

## Acknowledgements

JBA is the receipt of the studentship awards from the Health Data Research UK-The Alan Turing Institute Wellcome PhD Programme in Health Data Science (Grant Ref: 218529/Z/19/Z). We acknowledge support of the UKRI AI programme, and the Engineering and Physical Sciences Research Council, for CHAI - Causality in Healthcare AI Hub [grant number EP/Y028856/1]. The research was carried out at the National Institute for Health and Care Research (NIHR) Manchester Biomedical Research Centre (BRC) (NIHR203308). This research has been conducted using the UK Biobank resource under application number 101874.

Figure 1: (a) Directed acyclic graph of disease progression and diagnosis over $T$ timepoints. Individuals transition from no disease ($S_{n,t} = 0$) to early stage ($S_{n,t} = 1$) and late stage ($S_{n,t} = 2$). Disease stage $S_{n,t}$ and the observability attribute $A_n$, through its presentation $\tilde{A}_n$, influence whether a test is performed and its result, $R_{n,t}$, and thus diagnosis $D_{n,t}$. (b) Directed acyclic graph of the counterfactual intervention where the presentation of the observability attribute is fixed to the reference group, $\tilde{A}_n = 0$. The counterfactual diagnosis $D_{n,t}^{\tilde{A}_n=0}$ represents whether an individual would have been diagnosed if their likelihood of testing were equal to that of the reference group.

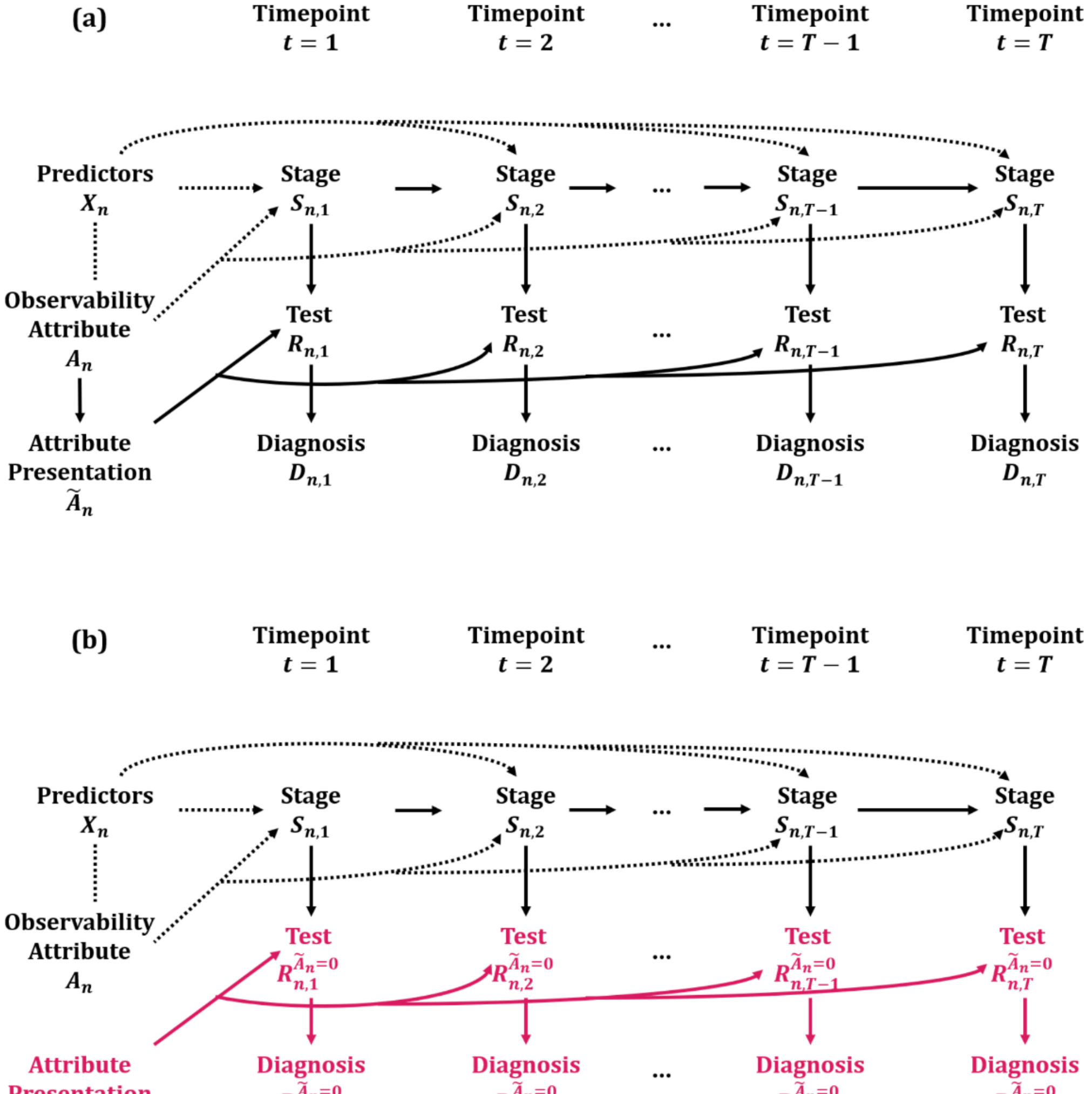

Table 1: Parameter estimates of the hidden Markov model across 200 simulation runs under Scenario 1, where all model assumptions are satisfied.

| **Parameter Name** | **True Value** | **Average Estimate** | **Bias** | **Empirical Standard Error** | **Mean Squared Error** |
|---|---|---|---|---|---|
| Testing rates and progression | | | | | |
| Testing rate $\beta^R_{S=0,A=0}$ | 0.025 | 0.025 | 0.000 | 0.000 | 0.0000 |
| Testing rate $\beta^R_{S=0,A=1}$ | 0.010 | 0.010 | 0.000 | 0.000 | 0.0000 |
| Testing rate $\beta^R_{S=1,A=0}$ | 0.100 | 0.102 | +0.002 | 0.030 | 0.0009 |
| Testing rate $\beta^R_{S=1,A=1}$ | 0.050 | 0.052 | +0.002 | 0.017 | 0.0003 |
| Testing rate $\beta^R_{S=2}$ | 0.300 | 0.302 | +0.002 | 0.122 | 0.0148 |
| Progression rate $\beta^{S=1\to 2}$ | 0.100 | 0.113 | +0.013 | 0.047 | 0.0024 |
| Hazard function | | | | | |
| Coefficient $\alpha_{X_1}$ | 0.500 | 0.503 | +0.003 | 0.026 | 0.0007 |
| Coefficient $\alpha_{X_2}$ | -0.250 | -0.254 | -0.004 | 0.030 | 0.0009 |
| Coefficient $\alpha_A$ | 0.250 | 0.244 | -0.006 | 0.091 | 0.0082 |
| Scale $\sigma$ | 0.005 | 0.005 | 0.000 | 0.001 | 0.0000 |
| Shape $\kappa$ | 1.500 | 1.508 | +0.008 | 0.109 | 0.0119 |

Figure 2: Average decile calibration plots across 200 simulated datasets for the four scenarios and four predictive models, stratified by the reference group ($A = 0$) and the underdiagnosed group ($A = 1$). Error bars represent the standard deviation of observed and expected probabilities within each decile.

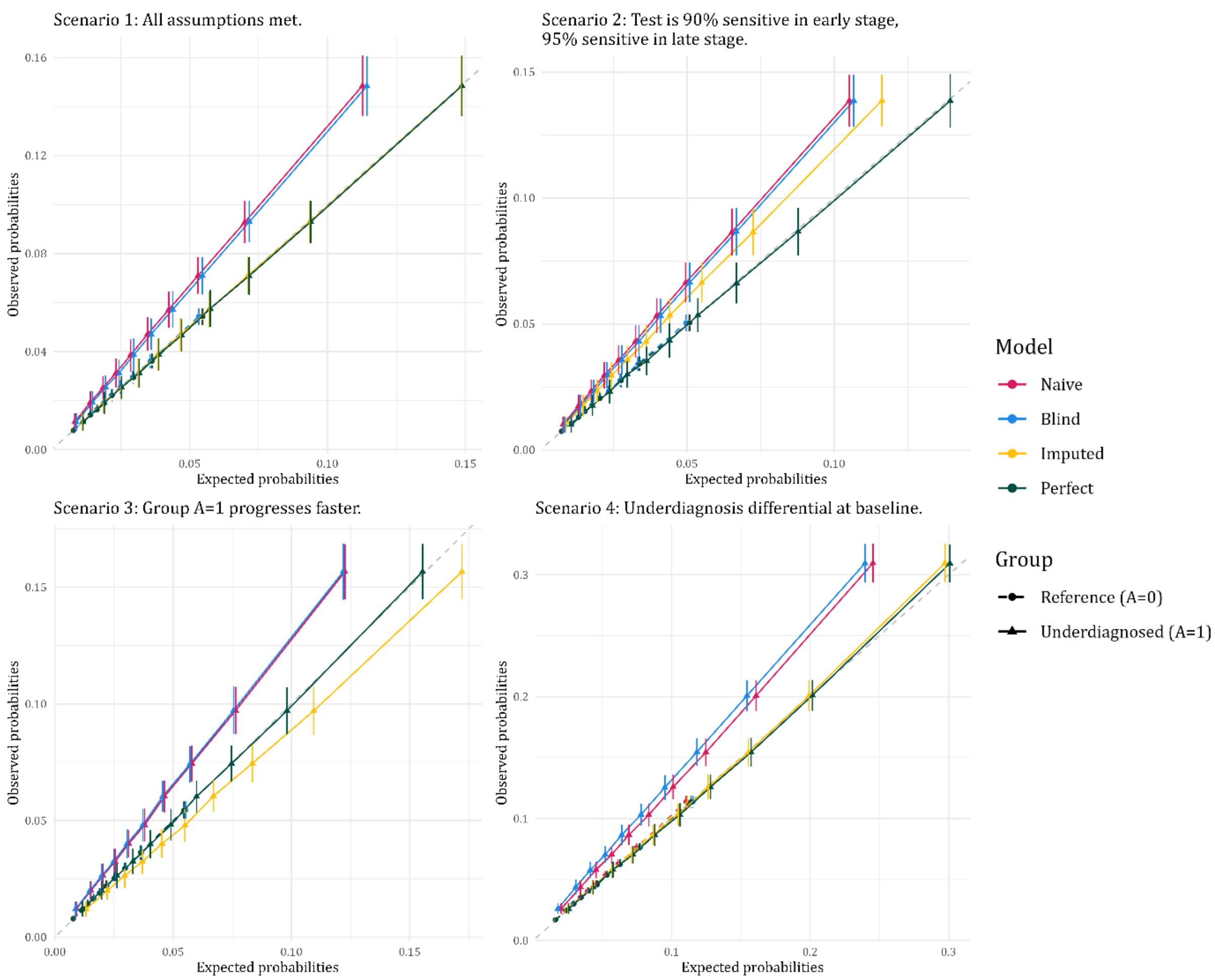

Table 2: Estimated hidden Markov model parameters for CKD prediction. 95% confidence intervals were obtained via bootstrapping.

| Parameter Description | Parameter Estimate (95% CI) | |
|---|---|---|
| **No CKD/Early CKD Testing Rate Model Coefficients** | | |
| | **Coefficient** | **Odds Ratio** |
| Age | 0.0386 (0.0365, 0.0405) | 1.04 (1.04, 1.04) |
| Diabetes | 2.34 (2.28, 2.40) | 10.36 (9.80, 11.02) |
| Hypertension | 1.02 (0.975, 1.07) | 2.78 (2.65, 2.92) |
| Cardiovascular Disease | 0.186 (0.123, 0.251) | 1.20 (1.13, 1.28) |
| Female | -0.364 (-0.400, -0.325) | 0.69 (0.67, 0.72) |
| In Most Deprived Townsend Quintile | 0.108 (0.058, 0.154) | 1.11 (1.06, 1.17) |
| Non-White | 0.397 (0.334, 0.489) | 1.49 (1.40, 1.63) |
| Intercept | -6.45 (-6.56, -6.33) | |
| Average No/Early CKD Testing Rate in the Population | 0.0252 (0.0249, 0.0257) | |
| **Other Coefficients** | | |
| Late CKD Testing Rate | 0.584 (0.564, 0.604) | |
| Early-to-Late Progression Rate | 0.125 (0.114, 0.139) | |
| Proportion of Late Stage Cases at First Timepoint | 0.289 (0.276, 0.320) | |

Figure 3: Prevalence of chronic kidney disease over time, as estimated by the hidden Markov model among individuals undiagnosed at baseline. The figure shows, at each timepoint, the proportions of individuals diagnosed at early or late stage and those remaining undiagnosed. Results are presented for the overall population and stratified by diabetes status at baseline.

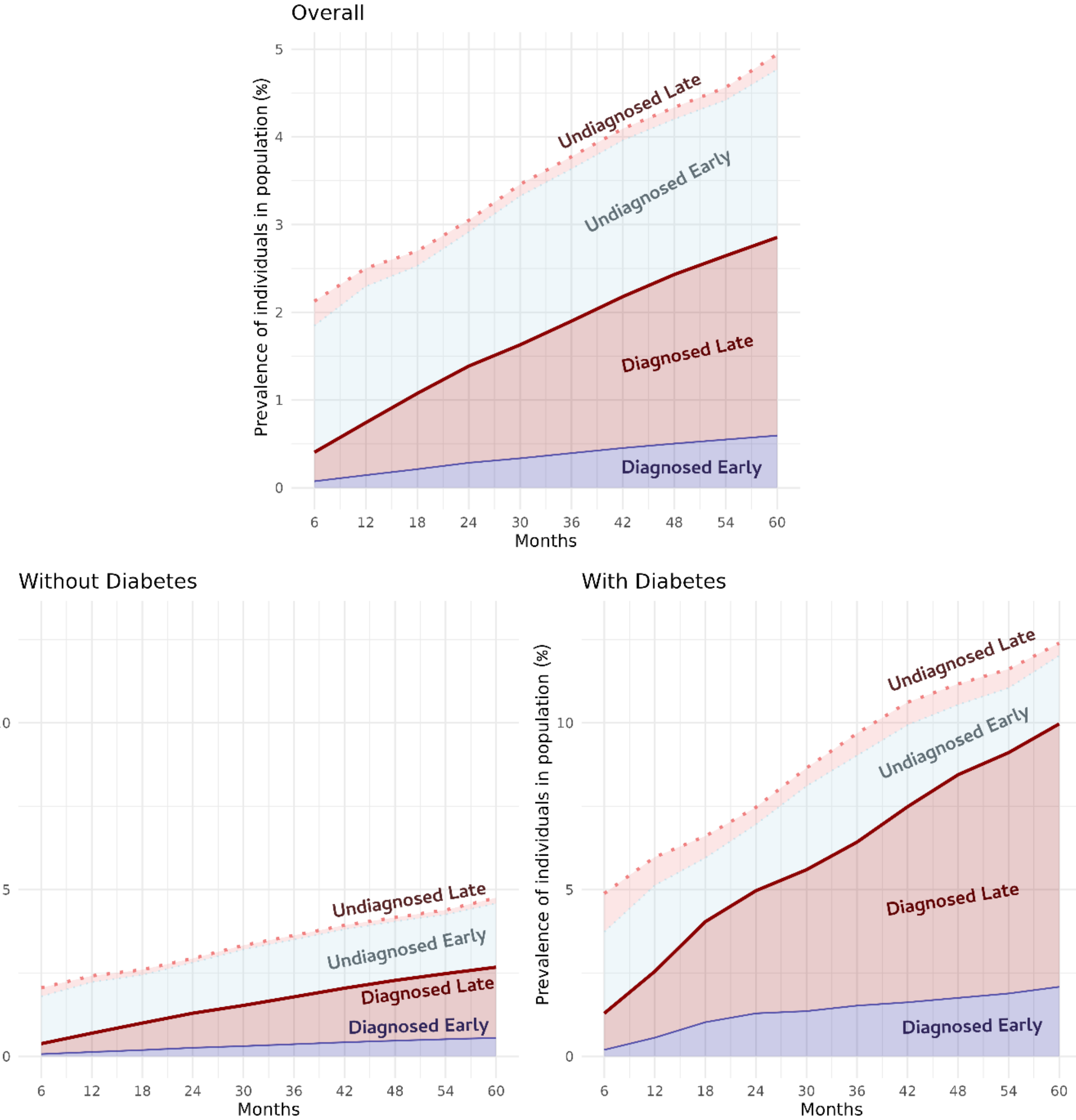

Figure 4: Quintile calibration plots for 5-year chronic kidney disease prediction using the proposed underdiagnosis-adjusted approach. Results are shown for the three models (Naïve, Blind, and Imputed) and are stratified by diabetes status at baseline.

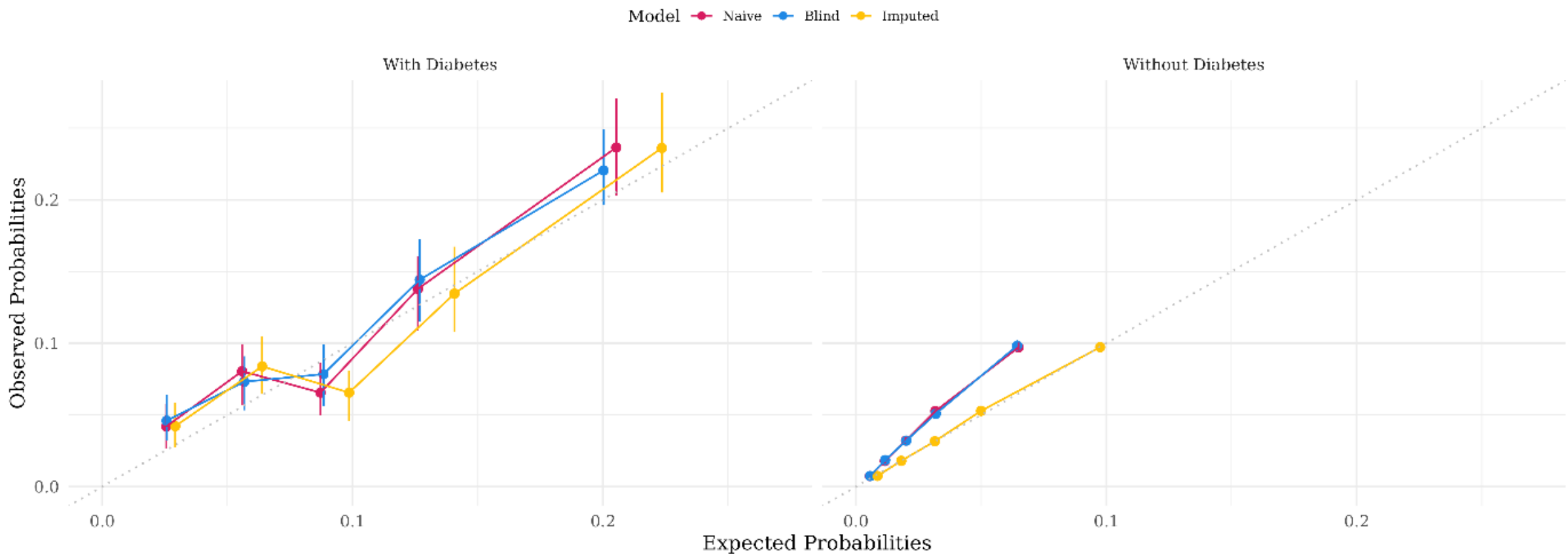

# Supplementary 1: Mathematical derivations for "Correcting heterogeneous diagnostic bias when developing clinical prediction models using causal hidden Markov models"

## Supplementary 1.1: Fitting the hidden Markov model through a maximum likelihood algorithm.

Consider a set of observations $\left\{x_n, a_n, \left\{r_{n,t}, d_{n,t}\right\}_{t=1..T_n}\right\}_{n=1..N}$ of patients entering a study with covariates $x_n$ and observability attribute $a_n$, which have tests $r_{n,t}$ and a diagnosis $d_{n,T_n} = 1$ if $r_{n,T_n} = 1$ *or* 2. Each individual is followed up for $T_n$ timepoints, where $T_n$ is, if they were diagnosed, the time at which the diagnosis happened (so $d_{n,T_n} = 1$), or, if they were not diagnosed, is the end of follow-up $T$.

The hidden Markov model follows the structure given in Figure 1 in the main text and is specified by its initial state $\pi(X_n, A_n)$, its transition kernel $\mathrm{Q}(X_n, A_n, t)$, and its emission probabilities $\Gamma(\mathrm{A_n})$, as defined in the main text:

$$\mathrm{Q_{ij}}(X_n, A_n, t) = P\left(S_{n,t} = j | X_n, A_n, S_{n,t-1} = i\right)$$
$$= \begin{pmatrix} 1 - h(X_n, A_n, t) & h(X_n, A_n, t) & 0 \\ 0 & 1 - \beta^{S=1\to2} & \beta^{S=1\to2} \\ 0 & 0 & 1 \end{pmatrix} \quad (S1.1)$$

$$\Gamma_{\mathrm{ij}}(A_n) = P\left(R_{n,t} = j \middle| A_n, S_{n,t} = i\right) = \begin{pmatrix} \beta^R_{S=0,A_n} & 0 & 0 & 1 - \beta^R_{S=0,A_n} \\ 0 & \beta^R_{S=1,A_n} & 0 & 1 - \beta^R_{S=1,A_n} \\ 0 & 0 & \beta^R_{S=2} & 1 - \beta^R_{S=2} \end{pmatrix} \quad (S1.2)$$

$$\pi_i(X_n, A_n) = P\left(S_{n,1} = i \middle| X_n, A_n\right) = \begin{pmatrix} 1 - h(X_n, A_n, 1) \\ (1 - \beta^{S_1=2})h(X_n, A_n, 1) \\ \beta^{S_1=2}h(X_n, A_n, 1) \end{pmatrix} \quad (S1.3)$$

At every timepoint, the probability of receiving $R_{n,t} = 0$ (test confirming no disease), $R_{n,t} = 1$ (test confirming early stage disease), $R_{n,t} = 2$ (test confirming late stage disease), and $R_{n,t} = 3$ (no tests received at this timepoint) depends on the current state $S_{n,t}$ and the emission matrix $\Gamma_{\mathrm{ij}}(A_n)$:

$$P\left(R_{n,t} = j | X_n, A_n\right) = \sum_{i=0,1,2} P\left(R_{n,t} = j \middle| X_n, A_n, S_{n,t} = i\right) \times P\left(S_{n,t} = i \middle| X_n, A_n\right)$$

$$= \sum_{i=0,1,2} \Gamma_{\mathrm{ij}}(A_n) \times P\left(S_{n,t} = i \middle| X_n, A_n\right)$$

$$= \begin{pmatrix} P(S_{n,t} = 0|X_n, A_n) \\ P(S_{n,t} = 1|X_n, A_n) \\ P(S_{n,t} = 2|X_n, A_n) \end{pmatrix} \begin{pmatrix} \beta^R_{S=0,A_n} & 0 & 0 & 1-\beta^R_{S=0,A_n} \\ 0 & \beta^R_{S=1,A_n} & 0 & 1-\beta^R_{S=1,A_n} \\ 0 & 0 & \beta^R_{S=2} & 1-\beta^R_{S=2} \end{pmatrix} \quad (S1.4)$$

Where $\beta^R_{S=0,A_n}$ ($\beta^R_{S=0,A_n=0}$ and $\beta^R_{S=0,A_n=1}$ depending on the observability attribute) is the testing rate in those without the disease, $\beta^R_{S=1,A_n}$ ($\beta^R_{S=1,A_n=0}$ and $\beta^R_{S=1,A_n=1}$ depending on the observability attribute) is the testing rate in those with early stage, and $\beta^R_{S=2}$ (independent of $A_n$) is the testing rate in those with late stage.

To fit this model, we use maximum likelihood, where we want to find the parameters $\hat{\theta} \in \Theta$ which maximise the probability of observing $\{x_n, a_n, r_{n,t}, d_{n,t}\}_{n=1..N,t=1..T_n}$:

$$\begin{aligned} \hat{\theta} &= argmax_{\theta\in\Theta} \prod_{n=1..N} P(R_{n,1..T_n} = r_{n,1..T_n}|X_n = x_n, A_n = a_n) \\ &= argmax_{\theta\in\Theta} \prod_{n=1..N} P(R_{n,T_n} = r_{n,T_n}|X_n = x_n, A_n = a_n, R_{n,1..T_n-1} = r_{n,1..T_n-1}) \\ &\qquad \times P(R_{n,1..T_n-1} = r_{n,1..T_n-1}|X_n = x_n, A_n = a_n) \\ &= argmax_{\theta\in\Theta} \prod_{n=1..N} \prod_{t=1..T_n} P(R_{n,t} = r_{n,t}|X_n = x_n, A_n = a_n, R_{n,1..t-1} = r_{n,1..t-1}) \quad (S1.5) \end{aligned}$$

Where $R_{n,1..t}$ is a short-hand expression for $R_{n,1}, R_{n,2}, \dots R_{n,t}$. We thus decompose the likelihood of observing every timepoint's test results, $P(R_{n,1..T_n} = r_{n,1..T_n}|X_n = x_n, A_n = a_n)$, into a product of every timepoint's test results given previous tests, $P(R_{n,t} = r_{n,t}|X_n = x_n, A_n = a_n, R_{n,1..t-1} = r_{n,1..t-1})$. Taking the log-likelihood for the maximisation algorithm instead:

$$\begin{aligned} \hat{\theta} = argmax_{\theta\in\Theta} \sum_{n=1..N} \sum_{t=1..T_n} \log\Big(P(R_{n,t} = r_{n,t}|X_n = x_n, A_n = a_n, R_{n,1..t-1} \\ = r_{n,1..t-1})\Big) \quad (S1.6) \end{aligned}$$

In order to maximise this, we need to, for every individual $n$ and timepoint $t$, find an expression of $P(R_{n,t}|X_n, A_n, R_{n,1..t-1})$ in terms of the parameters $\theta$ of the model. This can be done iteratively, following a lattice algorithm for hidden Markov models [1]. We show below that, at every timepoint, we can express $P(S_{n,t}|X_n, A_n, R_{n,1..t-1})$ and $P(R_{n,t}|X_n, A_n, R_{n,1..t-1})$ in terms of $\theta$.

For the first timepoint, this is trivial, as:

$$P(S_{n,1} = i|X_n, A_n) = \pi_i(X_n, A_n) \quad (S1.7)$$

And:

$$P(R_{n,1} = j|X_n, A_n) = \sum_{i=0,1,2} \Gamma_{ij}(A_n) \times \pi_i(X_n, A_n) \quad (S1.8)$$

Which follows from expressions $S1.1$ and $S1.4$. If for a timepoint $t$, we have expressed $P(S_{n,t}|X_n, A_n, R_{n,1..t-1})$ and $P(R_{n,t}|X_n, A_n, R_{n,1..t-1})$ in terms of the parameters $\theta$, we show that we can do the same for $P(S_{n,t+1}|X_n, A_n, R_{n,1..t})$ and $P(R_{n,t+1}|X_n, A_n, R_{n,1..t})$. First, we use Bayes' theorem to have an expression for the stage of an individual conditional on past *and present* tests, $P(S_{n,t}|X_n, A_n, R_{n,1..t})$:

$$\begin{aligned}
&P(S_{n,t} = i|X_n, A_n, R_{n,1..t-1}, R_{n,t} = j) \\
&= \frac{P(R_{n,t} = j|X_n, A_n, R_{n,1..t-1}, S_{n,t} = i)\, P(S_{n,t} = i|X_n, A_n, R_{n,1..t-1})}{P(R_{n,t} = j|X_n, A_n, R_{n,1..t-1})} \\
&= \frac{P(R_{n,t} = j|A_n, S_{n,t} = i)\, P(S_{n,t} = i|X_n, A_n, R_{n,1..t-1})}{P(R_{n,t} = j|X_n, A_n, R_{n,1..t-1})} \\
&= \Gamma_{\mathrm{ij}}(A_n) \frac{P(S_{n,t} = i|X_n, A_n, R_{n,1..t-1})}{P(R_{n,t} = j|X_n, A_n, R_{n,1..t-1})} \quad (S1.9)
\end{aligned}$$

Using expression $S1.4$. Following this, we can write the expression of $P(S_{n,t+1}|X_n, A_n, R_{n,1..t})$ as:

$$\begin{aligned}
&P(S_{n,t+1} = j|X_n, A_n, R_{n,1..t}) \\
&= \sum_{i=0,1,2} P(S_{n,t+1} = j|X_n, A_n, R_{n,1..t}, S_{n,t} = i) \times P(S_{n,t} = i|X_n, A_n, R_{n,1..t}) \\
&= \sum_{i=0,1,2} \mathrm{Q}_{ij}(X_n, A_n, t+1) \times P(S_{n,t} = i|X_n, A_n, R_{n,1..t}) \quad (S1.10)
\end{aligned}$$

Using $S1.3$ and our expression of $P(S_{n,t}|X_n, A_n, R_{n,1..t})$. Having now derived this, we can also find $P(R_{n,t+1}|X_n, A_n, R_{n,1..t})$ as:

$$\begin{aligned}
&P(R_{n,t+1} = j|X_n, A_n, R_{n,1..t}) \\
&= \sum_{i=0,1,2} P(R_{n,t+1} = j|X_n, A_n, R_{n,1..t}, S_{n,t+1} = i) P(S_{n,t+1} = i|X_n, A_n, R_{n,1..t}) \\
&= \sum_{i=0,1,2} \Gamma_{\mathrm{ij}}(A_n)\, P(S_{n,t+1} = i|X_n, A_n, R_{n,1..t}) \quad (S1.11)
\end{aligned}$$

Using $S1.4$. By recursion, we have thus shown how to write the log-likelihood $\sum_{n=1..N} \sum_{t=1..T_n} \log\left(P(R_{n,t} = r_{n,t}|X_n = x_n, A_n = a_n, R_{n,1..t-1} = r_{n,1..t-1})\right)$ in terms of the parameters $\theta$. The expressions above are differentiable, allowing the calculation of gradients using a package like PyTorch.

## Supplementary 1.2: Estimating the counterfactual risk of diagnosis.

The goal of this section is to get the probability $P(D_{n,T}^{\tilde{A}=0} = 1|X_n = x_n, A_n = a_n, R_{n,1..T_n} = r_{n,1..T_n})$, i.e., the probability of being diagnosed in the counterfactual world (where we intervene on $\tilde{A}$ as done in Figure 1.b of the manuscript), conditional on each individual's characteristics $x_n$ and $a_n$, and their testing history $r_{n,1..T_n}$. We will do this in two steps. First, we need to get our best guess probabilities of stage at each timepoint for each individual, $P(S_{n,t}|X_n, A_n, R_{n,1..T_n})$, and secondly, we will use this to estimate $P(D_{n,T}^{\tilde{A}=0} = 1|X_n = x_n, A_n = a_n, R_{n,1..T_n} = r_{n,1..T_n})$.

### Supplementary 1.2.a: Estimating $P(S_{n,t}|X_n, A_n, R_{n,1..T_n})$.

To get these estimates, we use a backwards algorithm, so that we first estimate $P(S_{n,T_n}|X_n, A_n, R_{n,1..T_n})$, and then show that one can get $P(S_{n,t}|X_n, A_n, R_{n,1..T_n})$ from $P(S_{n,t+1}|X_n, A_n, R_{n,1..T_n})$. First, for the last timepoint of an individual, if an individual has received a test, then the testing history necessarily tells us that they were at that stage, so $P(S_{n,T_n} = i|X_n, A_n, R_{n,1..T_n-1}, R_{n,T_n} = i) = 1, \forall i = \{0,1,2\}$. For those without a test ($r_{n,T_n} = 3$), we use the expression found in $S1.9$ using Bayes' theorem to get:

$$P(S_{n,T_n} = i|X_n, A_n, R_{n,1..T_n-1}, R_{n,T_n} = 3) = \Gamma_{i3}(A_n)\frac{P(S_{n,T_n} = i|X_n, A_n, R_{n,1..T_n-1})}{P(R_{n,T_n} = 3|X_n, A_n, R_{n,1..T_n-1})} \quad (S1.12)$$

We now show that, if we know $P(S_{n,t+1}|X_n, A_n, R_{n,1..T_n})$, we know $P(S_{n,t}|X_n, A_n, R_{n,1..T_n})$. We write the latter as:

$$P(S_{n,t} = j|X_n, A_n, R_{n,1..T_n}) = \sum_{i=0,1,2} P(S_{n,t} = j|X_n, A_n, R_{n,1..T_n}, S_{n,t+1} = i) \times P(S_{n,t+1} = i|X_n, A_n, R_{n,1..T_n}) \quad (S1.13)$$

Here, the second term $P(S_{n,t+1} = i|X_n, A_n, R_{n,1..T_n})$ has already been assumed to be estimated. We note that:

$$P(S_{n,t} = j|X_n, A_n, R_{n,1..T_n}, S_{n,t+1} = i) = P(S_{n,t} = j|X_n, A_n, R_{n,1..t}, S_{n,t+1} = i) \quad (S1.14)$$

As there is a conditional independence $S_{n,t} \perp R_{n,t+1..T_n}|S_{n,t+1}$ (if we know the next timepoint's stage, knowing future testing history does not give us any additional information). From this, we use Bayes' theorem to write:

$$P(S_{n,t} = j|X_n, A_n, R_{n,1..t}, S_{n,t+1} = i)$$

$$= P\left(S_{n,t+1} = i | X_n, A_n, R_{n,1..t}, S_{n,t} = j\right) \times \frac{P\left(S_{n,t} = j | X_n, A_n, R_{n,1..t}\right)}{P\left(S_{n,t+1} = i | X_n, A_n, R_{n,1..t}\right)}$$

$$= Q_{ji}(X_n, A_n, t+1) \times \frac{P\left(S_{n,t} = j | X_n, A_n, R_{n,1..t}\right)}{P\left(S_{n,t+1} = i | X_n, A_n, R_{n,1..t}\right)} \quad (S1.15)$$

Here, we have used the expression for $Q_{ji}(X_n, A_n, t+1)$ from $S1.3$. We have also derived $P\left(S_{n,t} = j | X_n, A_n, R_{n,1..t}\right)$ in $S1.9$ and $P\left(S_{n,t+1} = i | X_n, A_n, R_{n,1..t}\right)$ in $S1.10$. This gives us the expression:

$$P\left(S_{n,t} = j \middle| X_n, A_n, R_{n,1..T_n}\right)$$
$$= \sum_{i=0,1,2} Q_{ji}(X_n, A_n, t+1) \times \frac{P\left(S_{n,t} = j | X_n, A_n, R_{n,1..t}\right)}{P\left(S_{n,t+1} = i | X_n, A_n, R_{n,1..t}\right)}$$
$$\times P\left(S_{n,t+1} = i \middle| X_n, A_n, R_{n,1..T_n}\right) \quad (S1.16)$$

Supplementary 1.2.b: Estimating $P(D_{n,T}^{\tilde{A}=0} = 1 | X_n, A_n, R_{n,1..T_n})$.

Having now an expression for the stage of an individual at any timepoint conditioned on testing history, $P\left(S_{n,t} = j \middle| X_n, A_n, R_{n,1..T_n}\right)$, we will use this to compute the counterfactual probability of being diagnosed for this individual, $P\left(D_{n,T}^{\tilde{A}=0} = 1 \middle| X_n, A_n, R_{n,1..T_n}\right)$. We first note that this can again be calculated recursively, as:

$$P\left(D_{n,T}^{\tilde{A}=0} = 1 \middle| X_n, A_n, R_{n,1..T_n}\right)$$
$$= \sum_{t=1..T} P\left(D_{n,t}^{\tilde{A}=0} = 1 \middle| X_n, A_n, R_{n,1..T_n}, D_{n,t-1}^{\tilde{A}=0} = 0\right)$$
$$\times P\left(D_{n,t-1}^{\tilde{A}=0} = 0 \middle| X_n, A_n, R_{n,1..T_n}\right) \quad (S1.17)$$

For the first timepoint, the counterfactual probability of each test is:

$$P\left(R_{n,1}^{\tilde{A}=0} = j | X_n, A_n\right)$$
$$= \sum_{i=0,1,2} P\left(R_{n,1}^{\tilde{A}=0} = j | X_n, A_n, S_{n,1} = i\right) P\left(S_{n,1} = i | X_n, A_n\right)$$
$$= \sum_{i=0,1,2} \Gamma_{ij}^{\tilde{A}=0}(A_n) P\left(S_{n,1} = i | X_n, A_n\right) \ (S1.18)$$

Where we have already estimated $P\left(S_{n,1} = i | X_n, A_n\right)$, and $\Gamma_{\mathrm{ij}}^{\tilde{A}=0}(A_n) = \Gamma_{\mathrm{ij}}(0)$ is the counterfactual emission matrix. From this, $P\left(D_{n,1}^{\tilde{A}=0} = 1 | X_n, A_n\right) = P\left(R_{n,1}^{\tilde{A}=0} = 1 | X_n, A_n\right) + P\left(R_{n,1}^{\tilde{A}=0} = 2 | X_n, A_n\right)$.

Let's assume that we have determined the counterfactual probabilities $P\left(S_{n,t-1} \middle| X_n, A_n, R_{n,1..T_n}, D_{n,t-2}^{\tilde{A}=0} = 0\right)$, $P\left(R_{n,t-1}^{\tilde{A}=0} \middle| X_n, A_n, R_{n,1..T_n}, D_{n,t-2}^{\tilde{A}=0} = 0\right)$, and $P\left(D_{n,t-1}^{\tilde{A}=0} \middle| X_n, A_n, R_{n,1..T_n}, D_{n,t-2}^{\tilde{A}=0} = 0\right)$ for the previous timepoint $t-1$, and any previous

timepoints. We first find the probability of being at a particular stage, conditioned on not having been diagnosed before:

$$P\left(S_{n,t} = j \middle| X_n, A_n, R_{n,1..T_n}, D_{n,t-1}^{\tilde{A}=0} = 0\right)$$
$$= \sum_{i=0,1,2} P\left(S_{n,t} = j \middle| X_n, A_n, R_{n,1..T_n}, D_{n,t-1}^{\tilde{A}=0} = 0, S_{n,t-1} = i\right)$$
$$\times P\left(S_{n,t-1} = i \middle| X_n, A_n, R_{n,1..T_n}, D_{n,t-1}^{\tilde{A}=0} = 0\right) \quad (S1.19)$$

For the first term, $P\left(S_{n,t} = j \middle| X_n, A_n, R_{n,1..T_n}, D_{n,t-1}^{\tilde{A}=0} = 0, S_{n,t-1} = i\right)$, we first note that $S_{n,t} \perp D_{n,t-1}^{\tilde{A}=0} | S_{n,t-1}$, as conditioning on past stage introduces independence between current stage and any past observations. This gives, using Bayes' theorem twice:

$$P\left(S_{n,t} = j \middle| X_n, A_n, R_{n,1..T_n}, D_{n,t-1}^{\tilde{A}=0} = 0, S_{n,t-1} = i\right)$$
$$= P\left(S_{n,t} = j \middle| X_n, A_n, R_{n,1..T_n}, S_{n,t-1} = i\right)$$
$$= P\left(S_{n,t-1} = i \middle| X_n, A_n, R_{n,1..T_n}, S_{n,t} = j\right) \times \frac{P\left(S_{n,t} = j | X_n, A_n, R_{n,1..T_n}\right)}{P\left(S_{n,t-1} = i | X_n, A_n, R_{n,1..T_n}\right)}$$
$$= P\left(S_{n,t-1} = i \middle| X_n, A_n, R_{n,1..t-1}, S_{n,t} = j\right) \times \frac{P\left(S_{n,t} = j | X_n, A_n, R_{n,1..T_n}\right)}{P\left(S_{n,t-1} = i | X_n, A_n, R_{n,1..T_n}\right)}$$
$$= P\left(S_{n,t} = j \middle| X_n, A_n, R_{n,1..t-1}, S_{n,t-1} = i\right) \times \frac{P\left(S_{n,t-1} = i | X_n, A_n, R_{n,1..t-1}\right)}{P\left(S_{n,t} = j | X_n, A_n, R_{n,1..t-1}\right)}$$
$$\times \frac{P\left(S_{n,t} = j | X_n, A_n, R_{n,1..T_n}\right)}{P\left(S_{n,t-1} = i | X_n, A_n, R_{n,1..T_n}\right)}$$
$$= \mathrm{Q}_{ij}(X_n, A_n, t) \times \frac{P\left(S_{n,t-1} = i | X_n, A_n, R_{n,1..t-1}\right)}{P\left(S_{n,t} = j | X_n, A_n, R_{n,1..t-1}\right)} \times \frac{P\left(S_{n,t} = j | X_n, A_n, R_{n,1..T_n}\right)}{P\left(S_{n,t-1} = i | X_n, A_n, R_{n,1..T_n}\right)} \quad (S1.20)$$

Where between the second and third line, we have used $S_{n,t-1} \perp R_{n,t..T_n} | S_{n,t}$. These are all expressions we have already estimated. For the second expression of $S1.19$, we have:

$$P\left(S_{n,t-1} = i \middle| X_n, A_n, R_{n,1..T_n}, D_{n,t-1}^{\tilde{A}=0} = 0\right)$$
$$= P\left(D_{n,t-1}^{\tilde{A}=0} = 0 \middle| X_n, A_n, R_{n,1..T_n}, S_{n,t-1} = i\right) \times \frac{P\left(S_{n,t-1} = i \middle| X_n, A_n, R_{n,1..T_n}\right)}{P\left(D_{n,t-1}^{\tilde{A}=0} = 0 \middle| X_n, A_n, R_{n,1..T_n}\right)}$$
$$= \left(\Gamma_{i0}^{\tilde{A}=0}(A_n) + \Gamma_{i3}^{\tilde{A}=0}(A_n)\right) \times \frac{P\left(S_{n,t-1} = i \middle| X_n, A_n, R_{n,1..T_n}\right)}{P\left(D_{n,t-1}^{\tilde{A}=0} = 0 \middle| X_n, A_n, R_{n,1..T_n}\right)}$$
$$= \left(\Gamma_{i0}(0) + \Gamma_{i3}(0)\right) \times \frac{P\left(S_{n,t-1} = i \middle| X_n, A_n, R_{n,1..T_n}\right)}{P\left(D_{n,t-1}^{\tilde{A}=0} = 0 \middle| X_n, A_n, R_{n,1..T_n}\right)} \quad (S1.21)$$

We have already estimated these terms, so putting $S1.20$ and $S1.21$ gives us an expression in $S1.19$ for $P\left(S_t = j \middle| X_n, A_n, R_{1..T_n}, D_{n,t-1}^{\tilde{A}=0} = 0\right)$. For the testing rate:

$$P\left(R_{n,t}^{\tilde{A}=0} = j \middle| X_n, A_n, R_{n,1..T_n}, D_{n,t-1}^{\tilde{A}=0} = 0\right)$$

$$= \sum_{i=0,1,2} P\left(R_{n,t}^{\tilde{A}=0} = j \middle| X_n, A_n, R_{n,1..T_n}, D_{n,t-1}^{\tilde{A}=0} = 0, S_{n,t} = i\right)$$
$$\times P\left(S_{n,t} = i | X_n, A_n, R_{n,1..T_n}, D_{n,t-1}^{\tilde{A}=0} = 0\right)$$
$$= \sum_{i=0,1,2} \Gamma_{ij}(0) P\left(S_{n,t} = i | X_n, A_n, R_{n,1..T_n}, D_{n,t-1}^{\tilde{A}=0} = 0\right) \quad (S1.22)$$

And finally:

$$P\left(D_{n,t}^{\tilde{A}=0} = 1 \middle| X_n, A_n, R_{n,1..T_n}, D_{n,t-1}^{\tilde{A}=0} = 0\right)$$
$$= P\left(R_{n,t}^{\tilde{A}=0} = 1 \middle| X_n, A_n, R_{n,1..T_n}, D_{n,t-1}^{\tilde{A}=0} = 0\right)$$
$$+ P\left(R_{n,t}^{\tilde{A}=0} = 2 \middle| X_n, A_n, R_{n,1..T_n}, D_{n,t-1}^{\tilde{A}=0} = 0\right) \quad (S1.23)$$

We have thus estimated the probabilities of counterfactual diagnosis up to timepoint $T_n$. For timepoints beyond $T_n$, $t > T_n$, we can introduce the independence $S_{n,t} = j \perp R_{n,1..T_n}, D_{n,t-1}^{\tilde{A}=0} = 0 | S_{n,t-1} = i$, giving:

$$P\left(S_{n,t} = j \middle| X_n, A_n, R_{n,1..T_n}, D_{n,t-1}^{\tilde{A}=0} = 0, S_{n,t-1} = i\right)$$
$$= P\left(S_{n,t} = j \middle| X_n, A_n, S_{n,t-1} = i\right)$$
$$= \mathrm{Q}_{ij}(X_n, A_n) \quad (S1.24)$$

We can then use expressions $S1.22$ and $S1.23$ to get the testing and diagnosis rates. Thus, since we have, for all $1 \leq t \leq T$, derived $P\left(D_{n,t}^{\tilde{A}=0} = 1 \middle| X_n, A_n, R_{n,1..T_n}, D_{n,t-1}^{\tilde{A}=0} = 0\right)$, we can finally get an estimate for $P\left(D_{n,T}^{\tilde{A}=0} = 1 \middle| X_n, A_n, R_{n,1..T_n}\right)$ using expression $S1.17$.

# Supplementary 2: Details of the use of the proposed approach on the chronic kidney disease example.

## Supplementary 2.1: Data.

Data were obtained from UK Biobank, which recruited participants between 2006 and 2010 and provides demographic, clinical, and linked healthcare information. Linkage included both primary care records, to obtain testing history and stage-at-diagnosis, and inpatient hospital episode statistics, to further characterise stage-at-diagnosis. Participants without primary care linkage were excluded. Among those with linkage, only individuals registered with a TPP-managed practice at baseline were retained. Participants with a history of acute kidney injury (ICD-10: N17) or CKD (chronic kidney disease, ICD10: N18), present prior to or at baseline, were excluded. Censoring was only considered if it resulted directly from diagnosis, and follow-up ended at the time of CKD diagnosis or, for those without one, after 5 years from baseline.

## Supplementary 2.2: Predictors.

Nine predictors were collected in total. Age, sex, Townsend score, and body mass index (BMI) were recorded at baseline. Smoking status, also collected at baseline, was dichotomised into ever-smokers (current or former) and never-smokers. Ethnicity was collapsed into two groups: white (British, Irish, white, or any other white background) and non-white (all other categories). Clinical conditions were defined by ICD-10 diagnoses prior to study entry: E10–E11 for diabetes, I10 for hypertension, and I20, I21, I25, I48, I50, I63, I64, I65, I67, and I73 for cardiovascular disease. The Townsend score was dichotomised as being either inside or outside the top quintile of highest deprivation. Predictors were partitioned into observability attributes (entering the emission model and the hazard model, age, diabetes, CVD, hypertension, ethnicity, sex, and deprivation level) and risk-only covariates (entering the hazard model only, BMI and smoking), as described in the main Methods section.

Missing values in predictors were imputed using the mice package in R, with all other covariates included in the predictor matrix. A single imputed dataset was generated, and imputation was performed before model development and validation.

## Supplementary 2.3: Outcome definition.

The outcome was the 5-year incidence of CKD, identified using ICD-10 code N18. The prediction horizon of 5 years was chosen as it was a realistic choice for real prediction model development which might be affected by differences in diagnostic delay. Stage at diagnosis was derived from both primary and secondary care sources. In primary care, staging was based on CTV3 code lists (stages 1–5) available at https://github.com/UoM-

Data-Science-Platforms/gm-sde/tree/master/shared/clinical-code-sets/conditions. In secondary care, ICD-10 codes N18.1–N18.5 were used to define stage.

Across follow-up, 7090 incident CKD cases were identified. Of these, 932 had no recorded stage at diagnosis. Individuals with missing stage at diagnosis were imputed using predictive mean matching based on baseline covariates.

## Supplementary 2.4: Identifying negative tests.

Negative tests were defined as albumin-to-creatinine ratio (ACR) measurements with a measurement under 3 mg/mmol. ACR tests were identified from primary care records using the CTV3 code "Urine albumin/creatinine ratio." Tests conducted within six months prior to diagnosis were excluded, as these were considered part of the diagnostic process. All ACR tests between baseline and (if present) diagnosis were included.

## Supplementary 2.5: Fitting the hidden Markov model.

Data were structured into tensors for model fitting. Time was discretised into six-month intervals, with a maximum follow-up of 5 years, giving 10 timepoints per individual. Each patient record was encoded as a tensor of size (N, 10), where values indicated: 0, at least one ACR test in the interval confirming no CKD; 1, confirmed early stage CKD; 2, confirmed late stage CKD; 3, neither test nor diagnosis.

The HMM and its fitting process were implemented in Python using PyTorch. Transition and emission probabilities were restricted to the interval (0, 1). Late stage testing rate, $\beta^R_{S=2}$ was constrained to be greater than or equal to the maximum early-stage testing rate, $\max_{a_n} b(A_n = a_n)$.

Parameters were initialised arbitrarily (specific values provided in the code) and optimised using the Broyden–Fletcher–Goldfarb–Shanno algorithm, with a learning rate of 0.1 and up to 10000 iterations per step.

## Supplementary 2.6: Model fitting and validation.

The HMM was fitted to the full dataset, and counterfactual outcomes for 5-year incidence, $d^{cf}_{n,10}$, were generated at timepoint $t = 10$. Three logistic regression models were then developed:

1) A logistic regression 'naïve' model, using covariates and the observability attributes to predict observed 5-year incidence $d_{n,10}$.

2) A logistic regression 'blind' model, excluding ethnicity, sex, and Townsend score when predicting $d_{n,10}$.
3) A logistic regression 'imputed' model using covariates and the observability attributes to predict $d_{n,10}^{cf}$.

Model performance was first evaluated on the training data. To correct for optimism, we applied a bootstrap pipeline. For each iteration:

1) A sample of equal size to the original dataset was drawn with replacement.
2) The HMM was re-fitted in the bootstrapped sample. To improve computational efficiency, the parameters are initialised to those of the fit in the overall data.
3) Counterfactual outcomes are re-imputed, and the three logistic regression models re-trained in the bootstrap sample.
4) Performance metrics are calculated in the bootstrapped sample (optimistic) and the original data (realistic) and recorded.

This process was repeated 100 times, and the average difference between bootstrap and original performance was used to correct the apparent performance measures obtained from the full dataset.

## Supplementary 3: Additional results of the validatory simulation.

Supplementary 3, Table 1: Performance of parameter estimation of the true hidden Markov model across 200 runs of a simulation for Scenario 2, where the test has a sensitivity of 90% in early stage cases, and of 95% in late stage cases.

| Parameter Name | True Value | Average Estimate | Bias | Empirical Standard Error | Mean Squared Error |
|---|---|---|---|---|---|
| Testing rates and progression | | | | | |
| Testing rate $\beta^{R}_{S=0,A=0}$ | 0.025 | 0.025 | 0.000 | 0.000 | 0.0000 |
| Testing rate $\beta^{R}_{S=0,A=1}$ | 0.010 | 0.010 | 0.000 | 0.000 | 0.0000 |
| Testing rate $\beta^{R}_{S=1,A=0}$ | 0.100 | 0.440 | +0.340 | 0.116 | 0.1289 |
| Testing rate $\beta^{R}_{S=1,A=1}$ | 0.050 | 0.280 | +0.230 | 0.095 | 0.0618 |
| Testing rate $\beta^{R}_{S=2}$ | 0.300 | 0.536 | +0.236 | 0.159 | 0.0806 |
| Progression rate $\beta^{S=1\rightarrow 2}$ | 0.100 | 0.650 | +0.550 | 0.263 | 0.3712 |
| Hazard function | | | | | |
| Coefficient $\alpha_{X_1}$ | 0.500 | 0.493 | -0.007 | 0.027 | 0.0008 |
| Coefficient $\alpha_{X_2}$ | -0.250 | -0.245 | +0.005 | 0.030 | 0.0009 |
| Coefficient $\alpha_{A}$ | 0.250 | 0.048 | -0.202 | 0.081 | 0.0475 |
| Scale $\sigma$ | 0.005 | 0.002 | -0.003 | 0.000 | 0.0000 |
| Shape $\kappa$ | 1.500 | 2.046 | +0.546 | 0.128 | 0.3146 |

Supplementary 3, Table 2: Performance of parameter estimation of the true hidden Markov model across 200 runs of a simulation for Scenario 3, where the two groups have different rates of disease progression.

| **Parameter Name** | **True Value** | **Average Estimate** | **Bias** | **Empirical Standard Error** | **Mean Squared Error** |
|---|---|---|---|---|---|
| Testing rates and progression | | | | | |
| Testing rate $\beta^R_{S=0,A=0}$ | 0.025 | 0.025 | -0.000 | 0.000 | 0.0000 |
| Testing rate $\beta^R_{S=0,A=1}$ | 0.010 | 0.010 | -0.000 | 0.000 | 0.0000 |
| Testing rate $\beta^R_{S=1,A=0}$ | 0.100 | 0.108 | +0.008 | 0.030 | 0.0009 |
| Testing rate $\beta^R_{S=1,A=1}$ | 0.050 | 0.043 | -0.007 | 0.013 | 0.0002 |
| Testing rate $\beta^R_{S=2}$ | 0.300 | 0.333 | +0.033 | 0.117 | 0.0147 |
| Progression rate $\beta^{S=1\to 2}_{A=0}$ | 0.100 | 0.112 | +0.012 | 0.043 | 0.0020 |
| Progression rate $\beta^{S=1\to 2}_{A=1}$ | 0.130 | 0.112 | -0.018 | 0.043 | 0.0022 |
| Hazard function | | | | | |
| Coefficient $\alpha_{X_1}$ | 0.500 | 0.505 | +0.005 | 0.023 | 0.0005 |
| Coefficient $\alpha_{X_2}$ | -0.250 | -0.251 | -0.001 | 0.024 | 0.0006 |
| Coefficient $\alpha_A$ | 0.250 | 0.414 | +0.164 | 0.092 | 0.0354 |
| Scale $\sigma$ | 0.005 | 0.005 | 0.000 | 0.001 | 0.0000 |
| Shape $\kappa$ | 1.500 | 1.514 | +0.014 | 0.092 | 0.0086 |

Supplementary 3, Table 3: Performance of parameter estimation of the true hidden Markov model across 200 runs of a simulation for Scenario 4, where there is an underdiagnosis differential at baseline.

| **Parameter Name** | **True Value** | **Average Estimate** | **Bias** | **Empirical Standard Error** | **Mean Squared Error** |
|---|---|---|---|---|---|
| Testing rates and progression | | | | | |
| Testing rate $\beta^R_{S=0,A=0}$ | 0.025 | 0.025 | 0.000 | 0.000 | 0.0000 |
| Testing rate $\beta^R_{S=0,A=1}$ | 0.010 | 0.010 | 0.000 | 0.000 | 0.0000 |
| Testing rate $\beta^R_{S=1,A=0}$ | 0.100 | 0.082 | -0.018 | 0.012 | 0.0005 |
| Testing rate $\beta^R_{S=1,A=1}$ | 0.050 | 0.040 | -0.010 | 0.006 | 0.0001 |
| Testing rate $\beta^R_{S=2}$ | 0.300 | 0.417 | +0.117 | 0.109 | 0.0253 |
| Progression rate $\beta^{S=1\to 2}$ | 0.100 | 0.074 | -0.026 | 0.011 | 0.0008 |
| Hazard function | | | | | |
| Coefficient $\alpha_{X_1}$ | 0.500 | 0.498 | -0.002 | 0.015 | 0.0002 |
| Coefficient $\alpha_{X_2}$ | -0.250 | -0.250 | -0.000 | 0.019 | 0.0004 |
| Coefficient $\alpha_A$ | 0.250 | 0.352 | +0.102 | 0.053 | 0.0131 |
| Scale $\sigma$ | 0.005 | 0.017 | +0.012 | 0.008 | 0.0002 |
| Shape $\kappa$ | 1.500 | 0.270 | -1.230 | 0.073 | 1.5179 |

Supplementary 3, Table 4: Performance metric averages (standard deviation in brackets), stratified by group, across 200 runs of the simulation for Scenario 1, where all assumptions made by the model are met.

| | **Reference Group (A=0)** | | | | **Underserved Group (A=1)** | | | |
|---|---|---|---|---|---|---|---|---|
| | **Naïve** | **Blind** | **Imputed** | **Perfect** | **Naïve** | **Blind** | **Imputed** | **Perfect** |
| AUROC | 0.65 (0.01) | 0.65 (0.01) | 0.65 (0.01) | 0.65 (0.01) | 0.70 (0.01) | 0.70 (0.01) | 0.70 (0.01) | 0.70 (0.01) |
| Calibration Slope | 1.00 (0.01) | 1.00 (0.01) | 1.00 (0.01) | 1.00 (0.01) | 0.90 (0.02) | 0.91 (0.02) | 1.00 (0.04) | 1.00 (0.02) |
| Calibration Intercept | 0.001 (0.049) | 0.014 (0.048) | 0.001 (0.049) | -0.002 (0.047) | 0.314 (0.072) | 0.282 (0.057) | 0.004 (0.100) | -0.005 (0.063) |
| Observed:Expected Ratio | 1.00 (0.05) | 1.01 (0.05) | 1.00 (0.05) | 1.00 (0.05) | 1.34 (0.09) | 1.30 (0.07) | 1.01 (0.09) | 1.00 (0.06) |
| Brier Score | 0.0229 (0.0008) | 0.0229 (0.0008) | 0.0229 (0.0008) | 0.0229 (0.0008) | 0.0499 (0.0020) | 0.0499 (0.0020) | 0.0497 (0.0019) | 0.0497 (0.0019) |
| Logistic Error | 0.108 (0.003) | 0.108 (0.003) | 0.108 (0.003) | 0.108 (0.003) | 0.199 (0.007) | 0.199 (0.007) | 0.197 (0.006) | 0.197 (0.006) |

Supplementary 3, Table 5: Performance metric averages (standard deviation in brackets), stratified by group, across 200 runs of the simulation for Scenario 2, where the test has a sensitivity of 90% in early stage cases, and of 95% in late stage cases.

| | **Reference Group (A=0)** | | | | **Underserved Group (A=1)** | | | |
|---|---|---|---|---|---|---|---|---|
| | **Naïve** | **Blind** | **Imputed** | **Perfect** | **Naïve** | **Blind** | **Imputed** | **Perfect** |
| AUROC | 0.65 (0.01) | 0.65 (0.01) | 0.65 (0.01) | 0.65 (0.01) | 0.71 (0.01) | 0.71 (0.01) | 0.71 (0.01) | 0.71 (0.01) |
| Calibration Slope | 1.00 (0.01) | 1.00 (0.01) | 1.00 (0.01) | 1.00 (0.01) | 0.90 (0.02) | 0.91 (0.02) | 0.94 (0.03) | 1.00 (0.02) |
| Calibration Intercept | -0.002 (0.049) | 0.009 (0.046) | -0.003 (0.049) | -0.003 (0.048) | 0.307 (0.069) | 0.276 (0.056) | 0.197 (0.084) | -0.011 (0.064) |
| Observed:Expected Ratio | 1.00 (0.05) | 1.01 (0.05) | 1.00 (0.05) | 1.00 (0.05) | 1.33 (0.09) | 1.30 (0.07) | 1.20 (0.09) | 0.99 (0.06) |
| Brier Score | 0.0214 (0.0007) | 0.0214 (0.0007) | 0.0214 (0.0007) | 0.0214 (0.0007) | 0.0467 (0.0018) | 0.0467 (0.0018) | 0.0466 (0.0018) | 0.0465 (0.0017) |
| Logistic Error | 0.102 (0.003) | 0.102 (0.003) | 0.102 (0.003) | 0.102 (0.003) | 0.189 (0.006) | 0.188 (0.006) | 0.188 (0.006) | 0.187 (0.006) |

Supplementary 3, Table 6: Performance metric averages (standard deviation in brackets), stratified by group, across 200 runs of the simulation for Scenario 3, where the two groups have different rates of disease progression.

| | **Reference Group (A=0)** | | | | **Underserved Group (A=1)** | | | |
|---|---|---|---|---|---|---|---|---|
| | **Naïve** | **Blind** | **Imputed** | **Perfect** | **Naïve** | **Blind** | **Imputed** | **Perfect** |
| AUROC | 0.65 (0.01) | 0.65 (0.01) | 0.65 (0.01) | 0.65 (0.01) | 0.71 (0.01) | 0.71 (0.01) | 0.71 (0.01) | 0.71 (0.01) |
| Calibration Slope | 1.00 (0.01) | 1.00 (0.01) | 1.00 (0.01) | 1.00 (0.01) | 0.91 (0.02) | 0.91 (0.02) | 1.05 (0.04) | 1.00 (0.02) |
| Calibration Intercept | -0.000 (0.047) | -0.008 (0.045) | -0.001 (0.047) | 0.005 (0.048) | 0.270 (0.065) | 0.285 (0.055) | -0.121 (0.091) | -0.002 (0.059) |
| Observed:Expected Ratio | 1.00 (0.05) | 0.99 (0.04) | 1.00 (0.05) | 1.01 (0.05) | 1.28 (0.08) | 1.30 (0.07) | 0.90 (0.07) | 1.00 (0.05) |
| Brier Score | 0.0229 (0.0007) | 0.0229 (0.0007) | 0.0229 (0.0007) | 0.0229 (0.0007) | 0.0519 (0.0020) | 0.0519 (0.0020) | 0.0518 (0.0019) | 0.0517 (0.0019) |
| Logistic Error | 0.108 (0.003) | 0.108 (0.003) | 0.108 (0.003) | 0.108 (0.003) | 0.205 (0.007) | 0.205 (0.007) | 0.203 (0.006) | 0.203 (0.006) |

Supplementary 3, Table 7: Performance metric averages (standard deviation in brackets), stratified by group, across 200 runs of the simulation for Scenario 4, where there is an underdiagnosis differential at baseline.

| | **Reference Group (A=0)** | | | | **Underserved Group (A=1)** | | | |
|---|---|---|---|---|---|---|---|---|
| | **Naïve** | **Blind** | **Imputed** | **Perfect** | **Naïve** | **Blind** | **Imputed** | **Perfect** |
| AUROC | 0.66 (0.01) | 0.66 (0.01) | 0.66 (0.01) | 0.66 (0.01) | 0.71 (0.01) | 0.71 (0.01) | 0.71 (0.01) | 0.71 (0.01) |
| Calibration Slope | 0.99 (0.01) | 1.00 (0.01) | 0.99 (0.01) | 1.00 (0.01) | 0.89 (0.02) | 0.86 (0.01) | 1.00 (0.03) | 1.00 (0.02) |
| Calibration Intercept | 0.028 (0.030) | -0.000 (0.029) | 0.028 (0.030) | 0.002 (0.033) | 0.266 (0.041) | 0.330 (0.034) | 0.013 (0.057) | -0.002 (0.045) |
| Observed:Expected Ratio | 1.03 (0.03) | 1.00 (0.03) | 1.03 (0.03) | 1.00 (0.03) | 1.25 (0.04) | 1.32 (0.04) | 1.01 (0.05) | 1.00 (0.04) |
| Brier Score | 0.0466 (0.0010) | 0.0466 (0.0009) | 0.0466 (0.0010) | 0.0466 (0.0009) | 0.0978 (0.0021) | 0.0981 (0.0021) | 0.0970 (0.0020) | 0.0970 (0.0020) |
| Logistic Error | 0.191 (0.003) | 0.191 (0.003) | 0.191 (0.003) | 0.191 (0.003) | 0.335 (0.006) | 0.337 (0.006) | 0.332 (0.005) | 0.332 (0.005) |

## Supplementary 4: Additional results of the CKD example.

Supplementary 4, Table 1: Baseline characteristics of the cohort, and CKD outcomes.

| | Total Number of Patients: N=123460 |
|---|---|
| Age, Mean (SD) | 56.6 (8.04) |
| Female | 67157 (54.4%) |
| Townsend Quintile | |
| 1st (Least Deprived) | 24773 (20.1%) |
| 2nd | 24639 (20.0%) |
| 3rd | 24721 (20.0%) |
| 4th | 24700 (20.0%) |
| 5th (Most Deprived) | 24627 (19.9%) |
| Non-White | 6899 (5.6%) |
| BMI, Mean (SD) | 27.4 (4.73) |
| Ever Smoker | 42549 (34.5%) |
| Diabetes | 3019 (2.4%) |
| Hypertension | 31380 (25.4%) |
| Cardiovascular Disease | 11137 (9.0%) |
| CKD Diagnosed by 5 Years | 3524 (2.9%) |
| Diagnosed at Stage 1 | 81 (0.1%) |
| Diagnosed at Stage 2 | 632 (0.5%) |
| Diagnosed at Stage 3 | 2611 (2.1%) |
| Diagnosed at Stage 4 | 66 (0.1%) |
| Diagnosed at Stage 5 | 39 (0.0%) |
| Missing Stage | 95 (0.1%) |
| No CKD by 5 Years | 119936 (97.1%) |

Supplementary 4, Table 2: Estimated parameters for the Cox Proportional Hazards model for CKD prediction. 95% confidence intervals were obtained via bootstrapping.

| Parameter Description | Parameter Estimate (95% CI) |
|---|---|
| Proportional Hazards Coefficients | |
| BMI | 0.0436 (0.0385, 0.0477) |
| Ever Smoker | -0.0299 (-0.0866, 0.0356) |
| Age | 0.0828 (0.0807, 0.0852) |
| Diabetes | 0.291 (0.269, 0.417) |
| Hypertension | 0.505 (0.455, 0.584) |
| Cardiovascular Disease | 0.407 (0.324, 0.447) |
| Female | 0.226 (0.172, 0.298) |
| In Most Deprived Townsend Quintile | 0.158 (0.0744, 0.204) |
| Non-White | 0.21 (0.0782, 0.233) |
| Base Hazard | |
| At Timepoint t=6 months | 3.08e-05 (2.79e-05, 3.12e-05) |
| At Timepoint t=12 months | 5.38e-06 (5.38e-06, 8.1e-06) |
| At Timepoint t=18 months | 2.81e-06 (2.8e-06, 3.5e-06) |
| At Timepoint t=24 months | 5.29e-06 (3.77e-06, 5.3e-06) |
| At Timepoint t=30 months | 6.17e-06 (4.91e-06, 6.33e-06) |
| At Timepoint t=36 months | 4.68e-06 (4.67e-06, 6.41e-06) |
| At Timepoint t=42 months | 4.73e-06 (4.71e-06, 6.03e-06) |
| At Timepoint t=48 months | 3.71e-06 (3.15e-06, 3.76e-06) |
| At Timepoint t=54 months | 3.53e-06 (2.91e-06, 3.54e-06) |
| At Timepoint t=60 months | 5.85e-06 (4.87e-06, 5.85e-06) |

Supplementary 4, Figure 1: Prevalence of chronic kidney disease estimated by our model in individuals who were undiagnosed at baseline, separated by those who have, at each timepoint, been diagnosed at early or late stage, or which remain undiagnosed at each of the two stages. This is presented stratified by hypertension and cardiovascular disease at baseline.

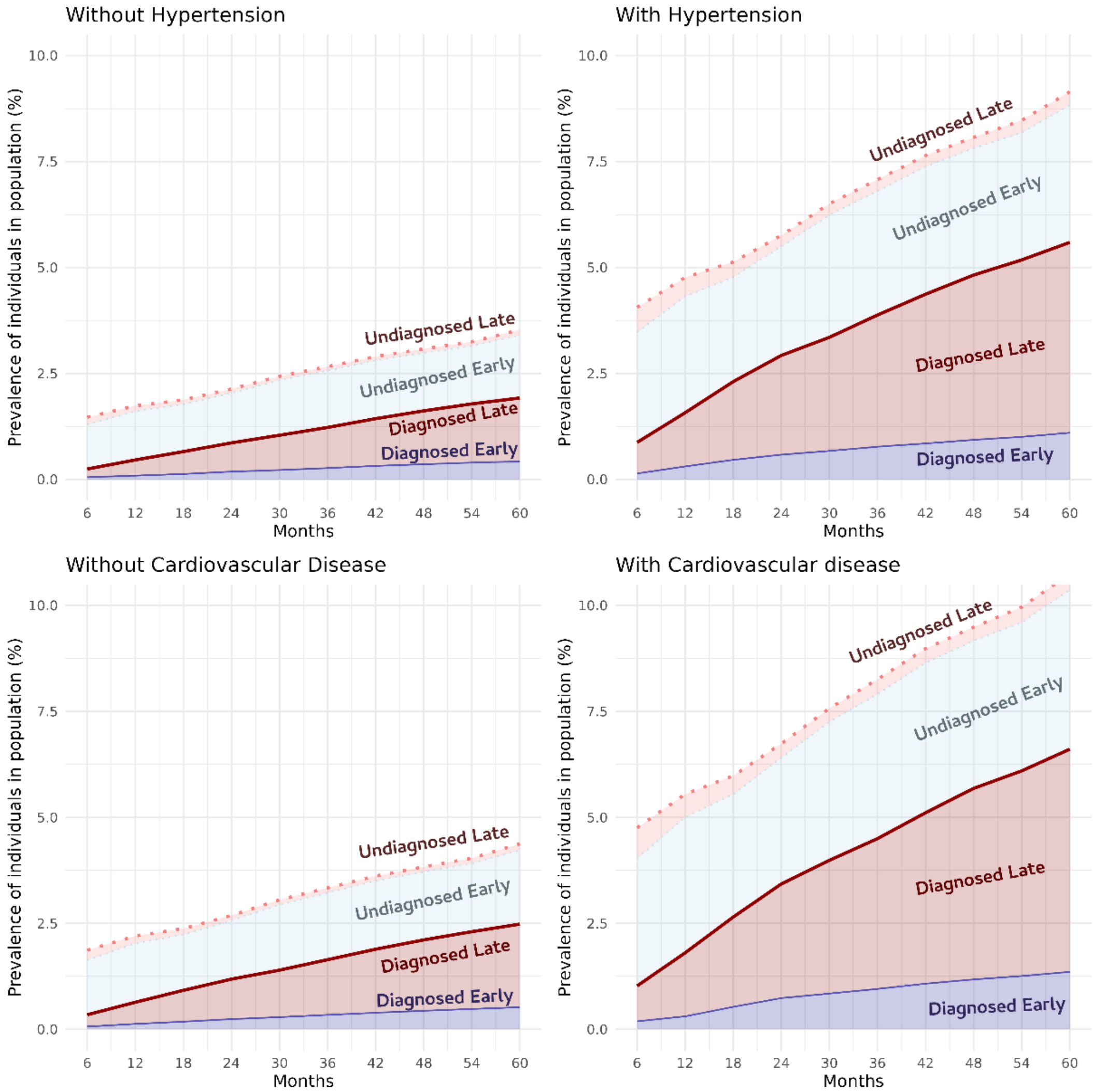

Supplementary 4, Figure 2: Prevalence of chronic kidney disease estimated by our model in individuals who were undiagnosed at baseline, separated by those who have, at each timepoint, been diagnosed at early or late stage, or which remain undiagnosed at each of the two stages. This is presented stratified by ethnicity, sex and whether they are in the top quintile of Townsend score of deprivation.

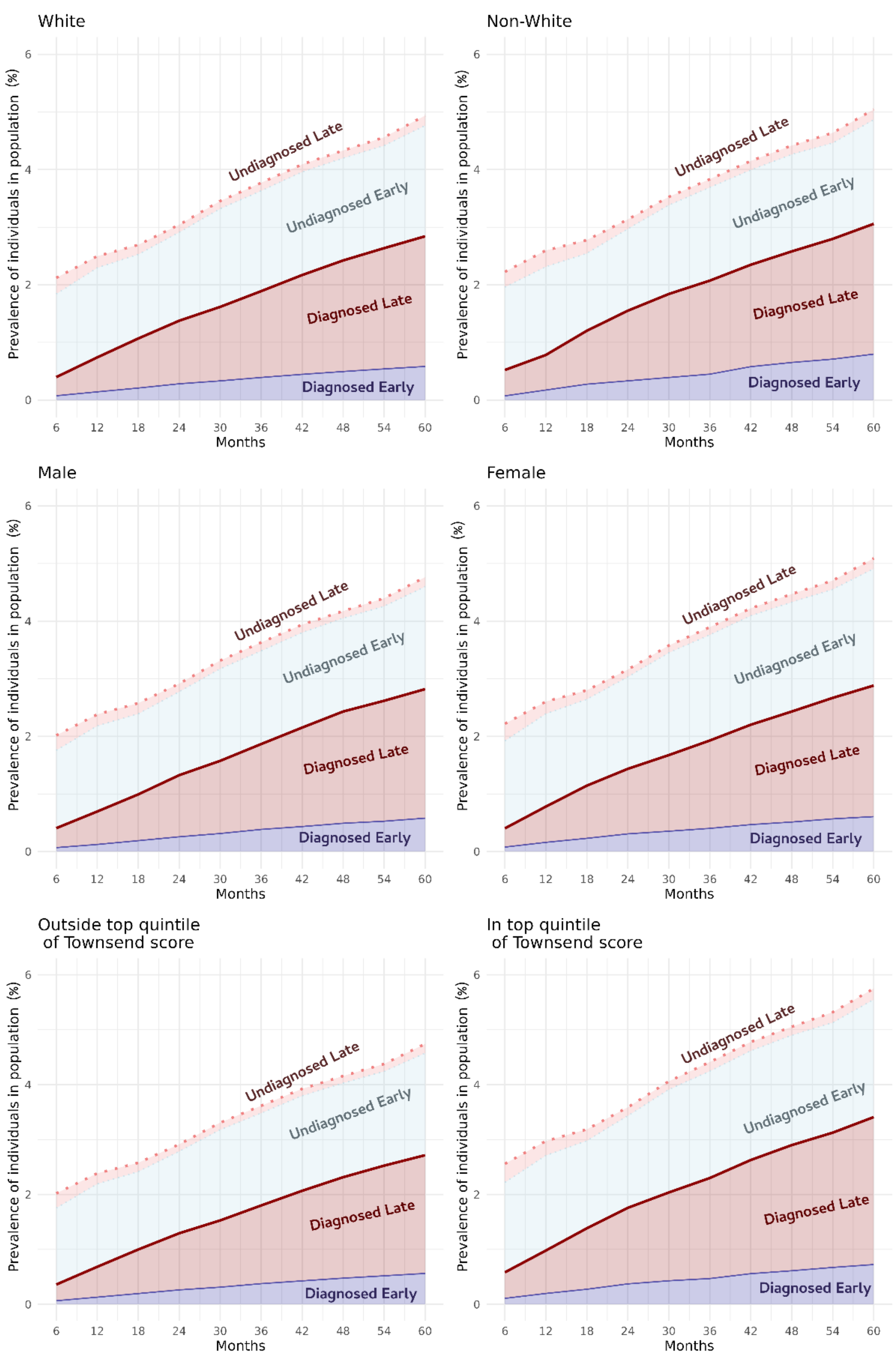

Supplementary 4, Figure 3: Quintile calibration plots for 5-year CKD prediction using the proposed approach to adjust for underdiagnosis for the three models considered, stratified by hypertension and cardiovascular disease at baseline.

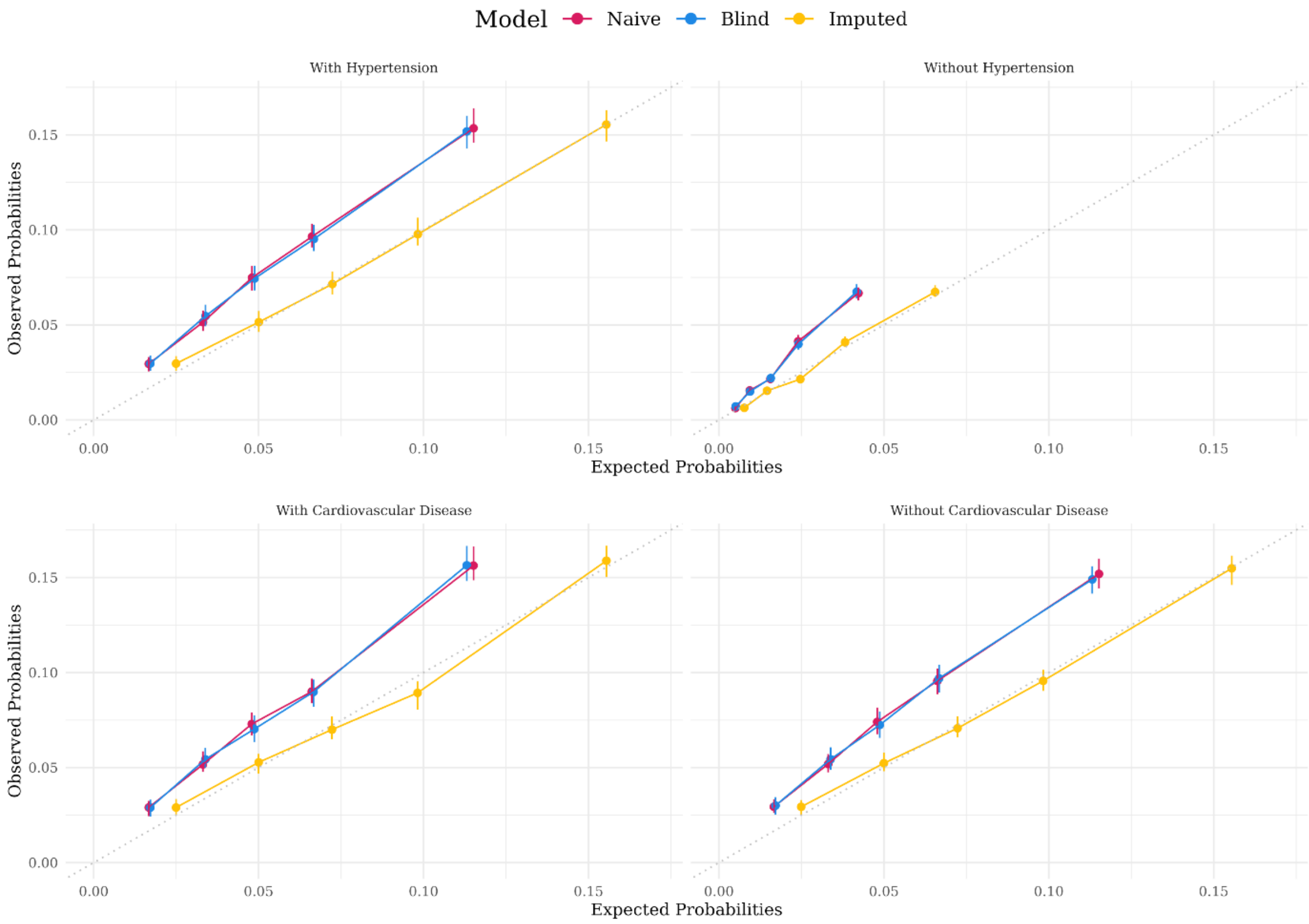

Supplementary 4, Figure 4: Quintile calibration plots for 5-year CKD prediction using the proposed approach to adjust for underdiagnosis for the three models considered, stratified by ethnicity, sex and whether they are in the top quintile of Townsend score of deprivation.

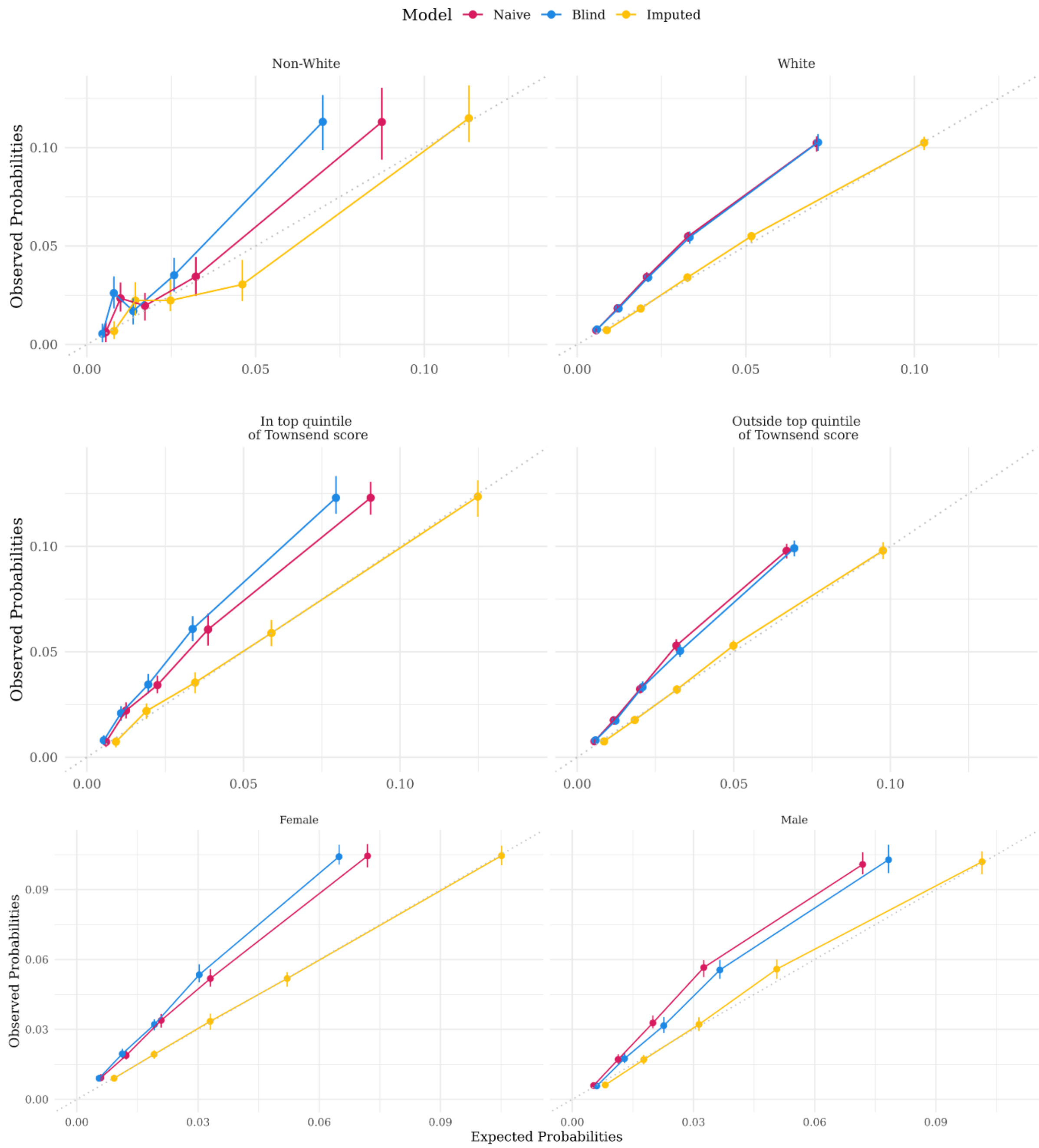

Supplementary 4, Table 3: Performance metrics (95% confidence intervals in brackets) of the three models in the overall population.

| | **Naïve** | **Blind** | **Imputed** |
|---|---|---|---|
| AUROC | 0.732 (0.727 - 0.739) | 0.731 (0.725 - 0.738) | 0.732 (0.727 - 0.739) |
| Calibration Slope | 0.869 (0.862 - 0.878) | 0.87 (0.861 - 0.877) | 0.997 (0.99 - 1.01) |
| Observed:Expected Ratio | 1.51 (1.48 - 1.55) | 1.51 (1.48 - 1.55) | 1.01 (0.981 - 1.03) |
| Logistic Error | 0.167 (0.163 - 0.171) | 0.167 (0.163 - 0.169) | 0.163 (0.160 - 0.166) |

Supplementary 4, Table 4: Performance metrics (95% confidence intervals in brackets), stratified by diabetes at baseline.

| | **Diabetes** | | | **No Diabetes** | | |
|---|---|---|---|---|---|---|
| | **Naïve** | **Blind** | **Imputed** | **Naïve** | **Blind** | **Imputed** |
| AUROC | 0.689 (0.657 - 0.724) | 0.684 (0.653 - 0.71) | 0.689 (0.657 - 0.72) | 0.728 (0.721 - 0.735) | 0.727 (0.719 - 0.733) | 0.728 (0.722 - 0.735) |
| Calibration Slope | 0.929 (0.87 - 0.97) | 0.926 (0.864 - 0.978) | 0.984 (0.925 - 1.06) | 0.868 (0.859 - 0.874) | 0.868 (0.861 - 0.877) | 0.998 (0.989 - 1.01) |
| Observed:Expected Ratio | 1.13 (1.03 - 1.24) | 1.13 (1.01 - 1.25) | 1.01 (0.91 - 1.09) | 1.55 (1.51 - 1.59) | 1.55 (1.52 - 1.59) | 1.01 (0.979 - 1.04) |
| Logistic Error | 0.33 (0.307 - 0.357) | 0.332 (0.314 - 0.354) | 0.33 (0.308 - 0.35) | 0.162 (0.159 - 0.165) | 0.163 (0.159 - 0.165) | 0.159 (0.157 - 0.162) |

Supplementary 4, Table 5: Performance metrics (95% confidence intervals in brackets), stratified by hypertension at baseline.

| | **Hypertension** | | | **No Hypertension** | | |
|---|---|---|---|---|---|---|
| | **Naïve** | **Blind** | **Imputed** | **Naïve** | **Blind** | **Imputed** |
| AUROC | 0.667 (0.658 - 0.68) | 0.662 (0.653 - 0.672) | 0.668 (0.659 - 0.679) | 0.712 (0.703 - 0.72) | 0.711 (0.701 - 0.719) | 0.712 (0.704 - 0.721) |
| Calibration Slope | 0.847 (0.834 - 0.863) | 0.848 (0.83 - 0.864) | 0.99 (0.973 - 1.01) | 0.879 (0.869 - 0.89) | 0.879 (0.872 - 0.89) | 1.00 (0.989 - 1.02) |
| Observed:Expected Ratio | 1.45 (1.41 - 1.49) | 1.45 (1.39 - 1.51) | 1.01 (0.974 - 1.04) | 1.57 (1.53 - 1.62) | 1.58 (1.52 - 1.64) | 1.01 (0.966 - 1.04) |
| Logistic Error | 0.274 (0.266 - 0.285) | 0.274 (0.265 - 0.285) | 0.268 (0.261 - 0.275) | 0.130 (0.125 - 0.134) | 0.130 (0.126 - 0.134) | 0.127 (0.124 - 0.131) |

Supplementary 4, Table 6: Performance metrics (95% confidence intervals in brackets), stratified by cardiovascular disease at baseline.

| | **Cardiovascular Disease** | | | **No Cardiovascular Disease** | | |
|---|---|---|---|---|---|---|
| | **Naïve** | **Blind** | **Imputed** | **Naïve** | **Blind** | **Imputed** |
| AUROC | 0.669 (0.659 - 0.681) | 0.667 (0.656 - 0.674) | 0.669 (0.659 - 0.679) | 0.666 (0.655 - 0.677) | 0.662 (0.654 - 0.671) | 0.667 (0.659 - 0.677) |
| Calibration Slope | 0.853 (0.838 - 0.866) | 0.855 (0.844 - 0.87) | 0.997 (0.979 - 1.01) | 0.849 (0.838 - 0.863) | 0.85 (0.836 - 0.868) | 0.992 (0.974 - 1.01) |
| Observed:Expected Ratio | 1.44 (1.38 - 1.48) | 1.43 (1.38 - 1.48) | 0.999 (0.967 - 1.04) | 1.45 (1.4 - 1.52) | 1.45 (1.39 - 1.5) | 1.01 (0.97 - 1.04) |
| Logistic Error | 0.27 (0.262 - 0.279) | 0.27 (0.264 - 0.278) | 0.265 (0.258 - 0.27) | 0.273 (0.266 - 0.28) | 0.273 (0.266 - 0.281) | 0.267 (0.259 - 0.275) |

Supplementary 4, Table 7: Performance metrics (95% confidence intervals in brackets), stratified by ethnicity.

| | White | | | Non-White | | |
|---|---|---|---|---|---|---|
| | **Naïve** | **Blind** | **Imputed** | **Naïve** | **Blind** | **Imputed** |
| AUROC | 0.731 (0.725 - 0.737) | 0.729 (0.724 - 0.736) | 0.731 (0.725 - 0.737) | 0.754 (0.723 - 0.78) | 0.753 (0.726 - 0.787) | 0.754 (0.729 - 0.784) |
| Calibration Slope | 0.867 (0.858 - 0.877) | 0.871 (0.864 - 0.88) | 0.997 (0.986 - 1) | 0.909 (0.881 - 0.949) | 0.852 (0.822 - 0.875) | 1.01 (0.981 - 1.06) |
| Observed:Expected Ratio | 1.53 (1.49 - 1.57) | 1.51 (1.47 - 1.56) | 1.01 (0.984 - 1.04) | 1.3 (1.17 - 1.46) | 1.61 (1.45 - 1.76) | 0.959 (0.84 - 1.07) |
| Logistic Error | 0.168 (0.164 - 0.171) | 0.168 (0.164 - 0.171) | 0.164 (0.161 - 0.167) | 0.15 (0.135 - 0.167) | 0.152 (0.139 - 0.171) | 0.149 (0.136 - 0.161) |

Supplementary 4, Table 8: Performance metrics (95% confidence intervals in brackets), stratified by sex.

| | Male | | | Female | | |
|---|---|---|---|---|---|---|
| | **Naïve** | **Blind** | **Imputed** | **Naïve** | **Blind** | **Imputed** |
| AUROC | 0.739 (0.731 - 0.75) | 0.739 (0.728 - 0.749) | 0.74 (0.73 - 0.748) | 0.727 (0.718 - 0.736) | 0.727 (0.718 - 0.735) | 0.727 (0.717 - 0.736) |
| Calibration Slope | 0.871 (0.862 - 0.882) | 0.901 (0.889 - 0.913) | 0.996 (0.983 - 1.01) | 0.868 (0.858 - 0.88) | 0.846 (0.836 - 0.853) | 0.999 (0.988 - 1.01) |
| Observed:Expected Ratio | 1.51 (1.45 - 1.58) | 1.36 (1.31 - 1.41) | 1.02 (0.984 - 1.06) | 1.52 (1.47 - 1.58) | 1.66 (1.61 - 1.71) | 1 (0.97 - 1.03) |
| Logistic Error | 0.164 (0.16 - 0.17) | 0.163 (0.157 - 0.169) | 0.161 (0.157 - 0.165) | 0.169 (0.164 - 0.173) | 0.17 (0.165 - 0.175) | 0.165 (0.161 - 0.169) |

Supplementary 4, Table 9: Performance metrics (95% confidence intervals in brackets), stratified by whether individuals are in the top quintile of Townsend score of deprivation.

| | **Outside Top Quintile of Townsend Score** | | | **Inside Top Quintile of Townsend Score** | | |
|---|---|---|---|---|---|---|
| | **Naïve** | **Blind** | **Imputed** | **Naïve** | **Blind** | **Imputed** |
| AUROC | 0.728 (0.72 - 0.735) | 0.727 (0.72 - 0.735) | 0.729 (0.719 - 0.734) | 0.744 (0.731 - 0.757) | 0.744 (0.73 - 0.758) | 0.744 (0.731 - 0.755) |
| Calibration Slope | 0.869 (0.86 - 0.878) | 0.879 (0.871 - 0.888) | 0.998 (0.987 - 1.01) | 0.872 (0.856 - 0.892) | 0.836 (0.822 - 0.857) | 0.997 (0.974 - 1.01) |
| Observed:Expected Ratio | 1.53 (1.49 - 1.59) | 1.48 (1.43 - 1.51) | 1.01 (0.983 - 1.04) | 1.45 (1.36 - 1.56) | 1.66 (1.56 - 1.76) | 1 (0.952 - 1.05) |
| Logistic Error | 0.163 (0.159 - 0.168) | 0.163 (0.159 - 0.167) | 0.159 (0.155 - 0.163) | 0.181 (0.174 - 0.19) | 0.184 (0.176 - 0.192) | 0.178 (0.17 - 0.186) |

Supplementary 4, Figure 5: Decision curve analysis of the three models, stratified by diabetes at baseline.

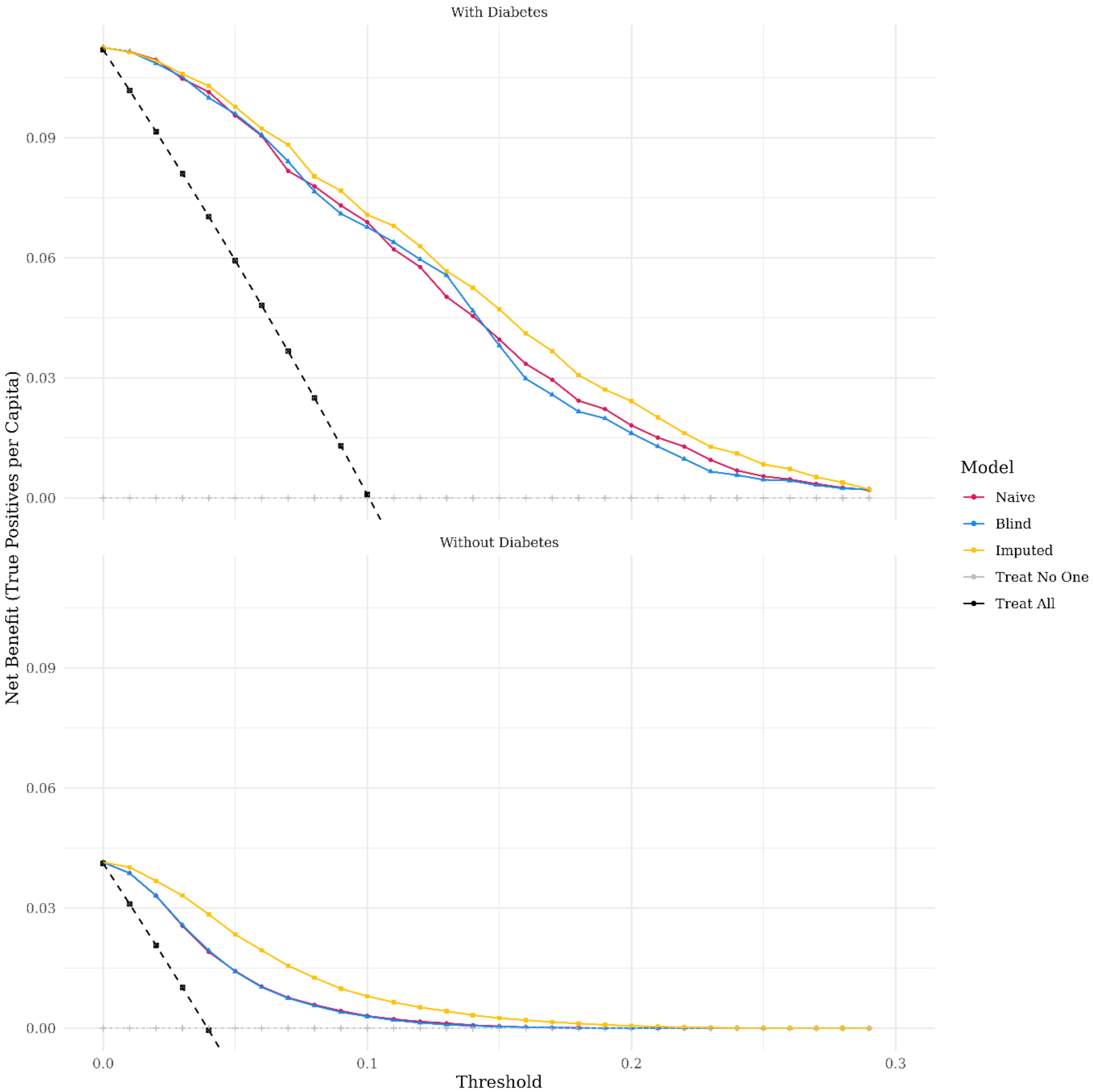

Supplementary 4, Figure 6: Decision curve analysis of the three models, stratified by hypertension at baseline.

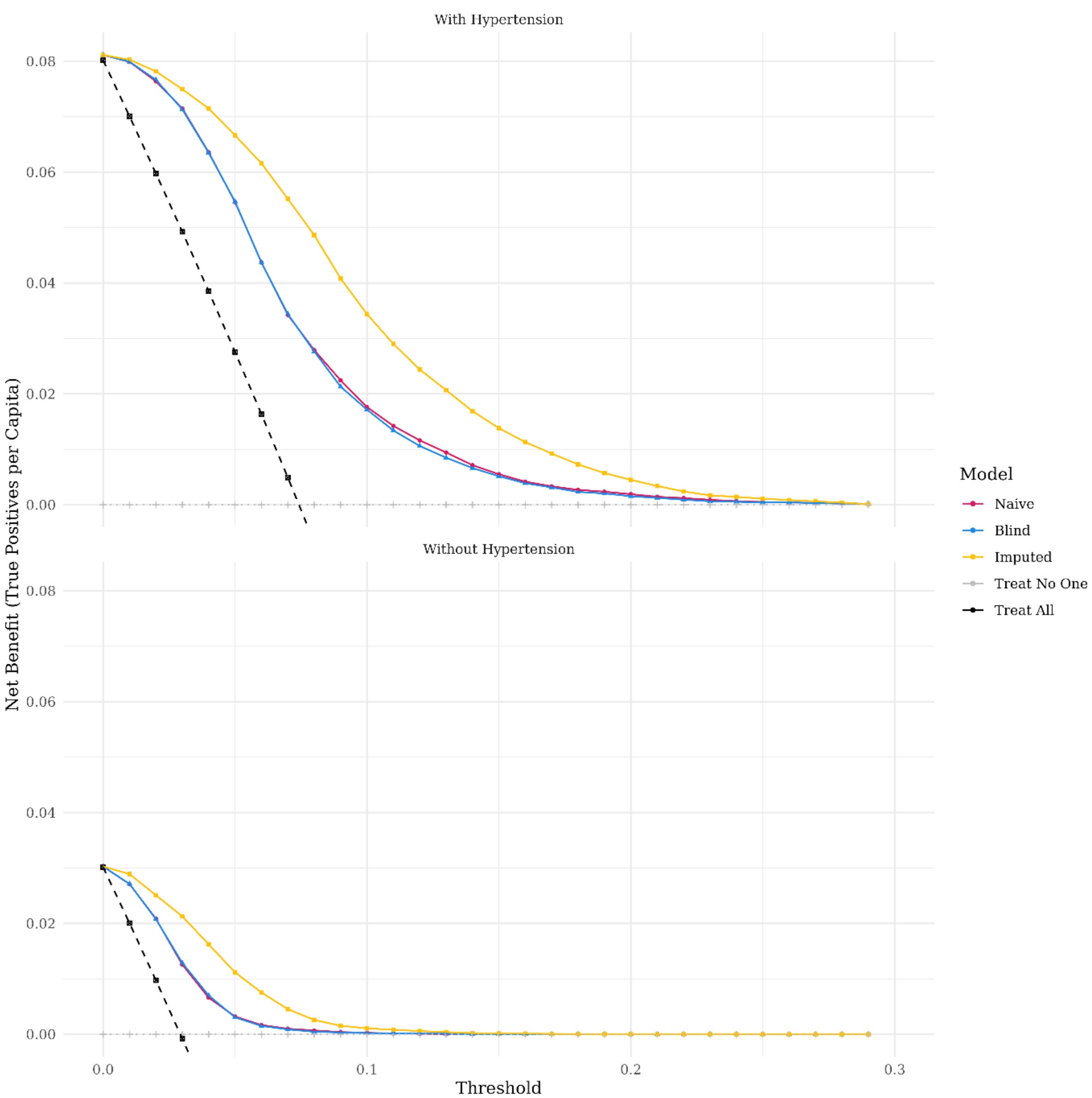

Supplementary 4, Figure 7: Decision curve analysis of the three models, stratified by cardiovascular disease at baseline.

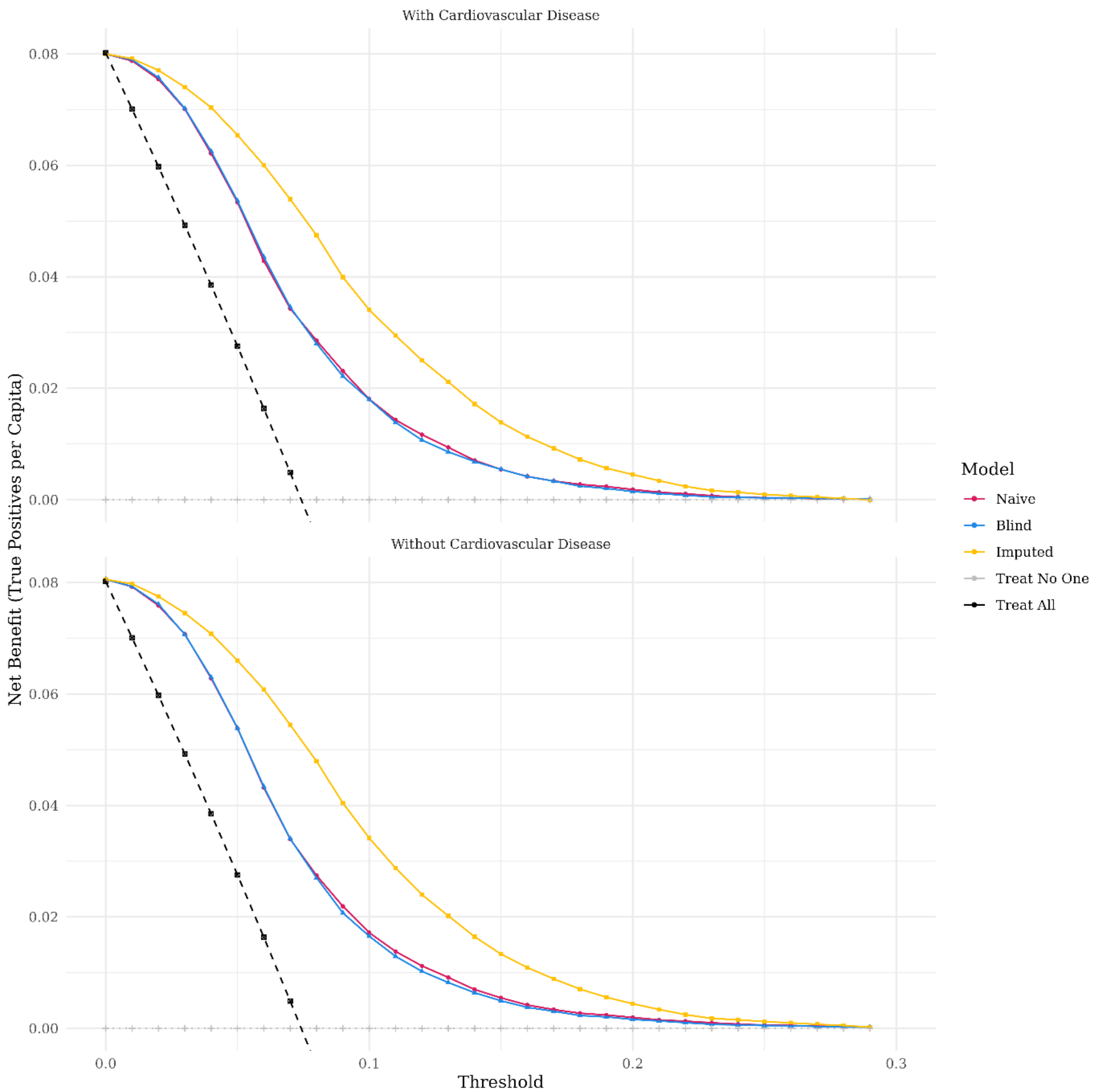

Supplementary 4, Figure 8: Decision curve analysis of the three models, stratified by ethnicity.

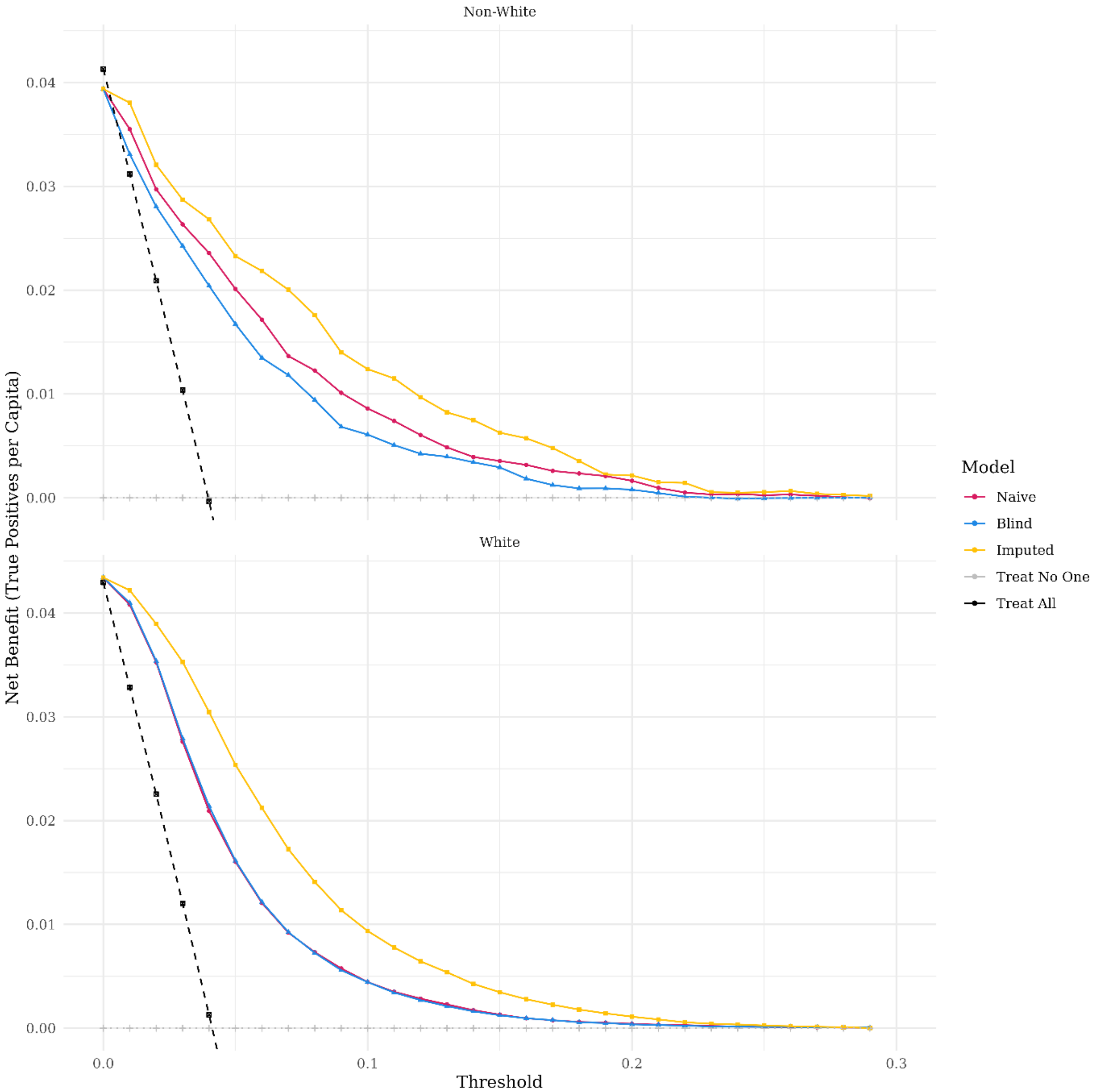

Supplementary 4, Figure 9: Decision curve analysis of the three models, stratified by sex.

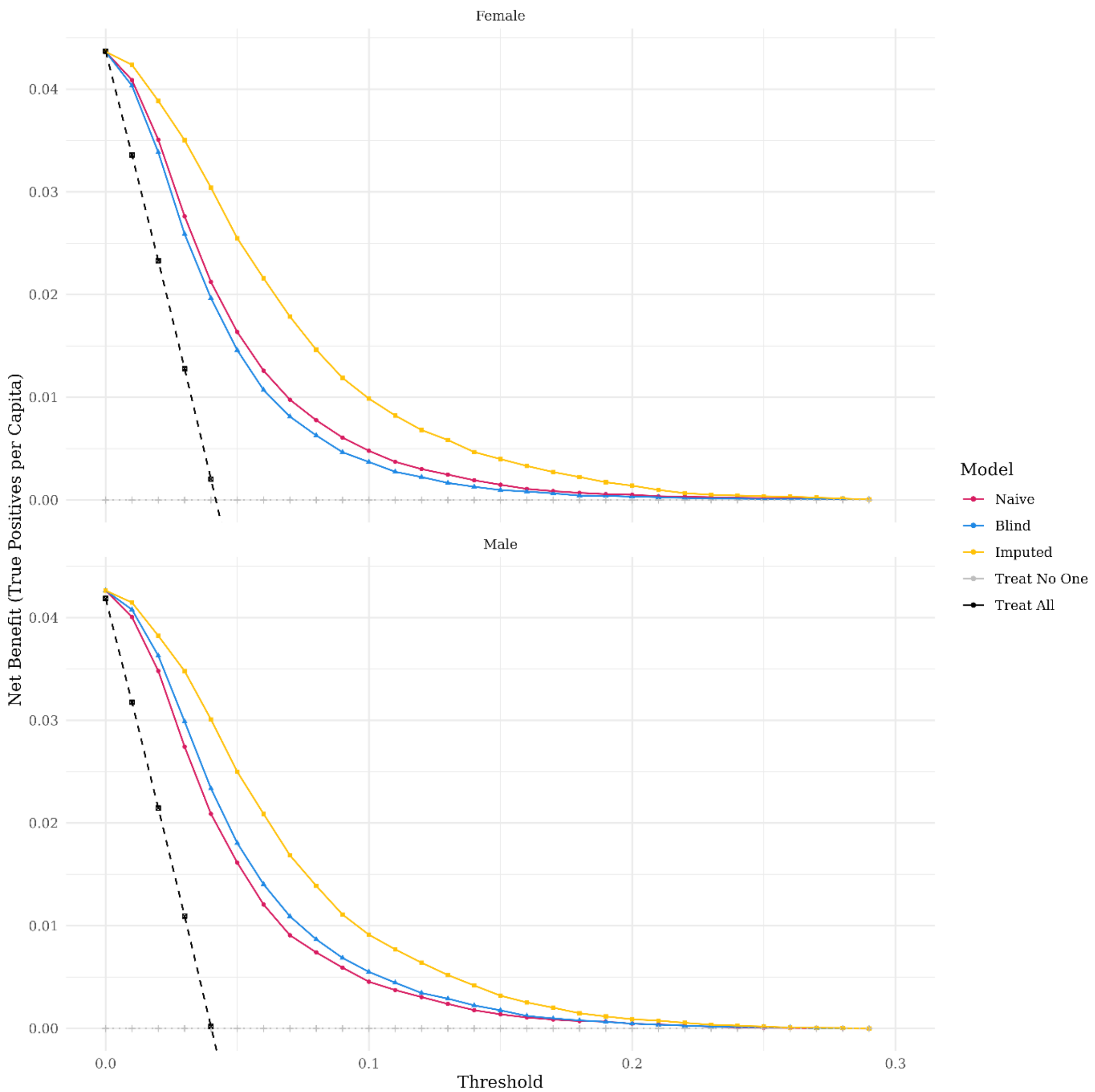

Supplementary 4, Figure 10: Decision curve analysis of the three models, stratified by whether individuals are in the top quintile of Townsend score of deprivation.

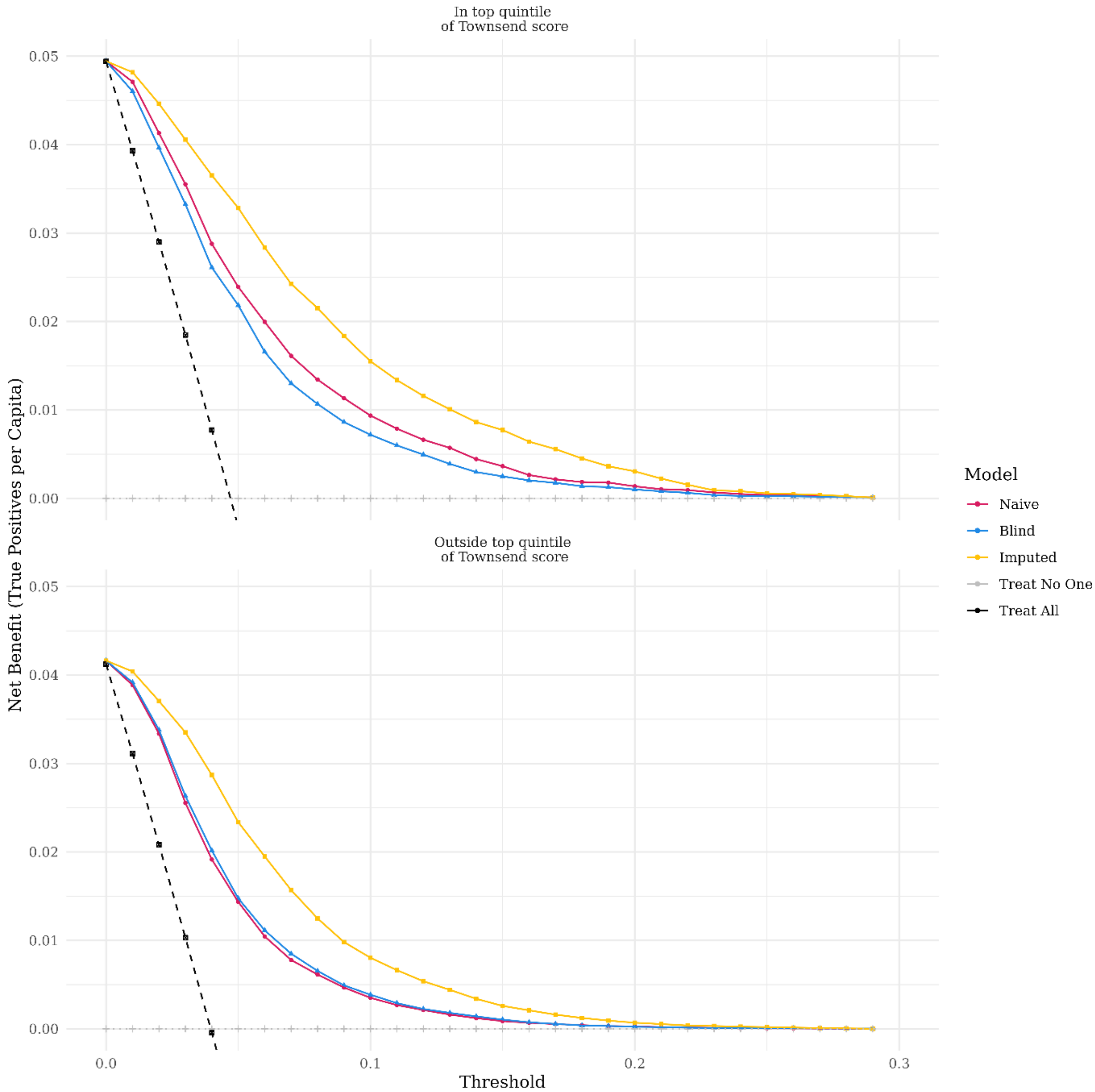

Supplementary 4, Table 10: Final coefficients of each of the clinical prediction models developed.

| ***Predictor Name*** | ***Naïve Model*** | ***Blind Model*** | ***Imputed Model*** |
|---:|---:|---:|---:|
| ***BMI*** | 0.05173 | 0.053518 | 0.048958 |
| ***Ever Smoker*** | -0.02199 | -0.06015 | -0.05125 |
| ***Age*** | 0.087895 | 0.086417 | 0.090976 |
| ***Diabetes*** | 0.683311 | 0.682482 | 0.389229 |
| ***Hypertension*** | 0.588849 | 0.58546 | 0.561386 |
| ***Cardiovascular Disease*** | 0.429006 | 0.408118 | 0.386749 |
| ***Female*** | 0.218717 | 0 | 0.219048 |
| ***In Top Quintile of Deprivation*** | 0.169069 | 0 | 0.157194 |
| ***Non-White*** | 0.215997 | 0 | 0.143665 |
| ***Intercept*** | -10.6674 | -10.4437 | -10.2819 |